\documentclass[twocolumn,10pt,aps,prb,notitlepage,showpacs,floatfix,unsortedaddress,superscriptaddress,citeautoscript]{revtex4-2}
\usepackage{amsmath,amssymb,amstext,graphicx,bm}
\usepackage{graphicx,color,bm}
\usepackage[english]{babel}
\usepackage{dsfont}
\usepackage{epsfig}
\newcommand{\vb}[1]{\bm{#1}} 

\begin{document}

\title{Anisotropic magnetoresistance of two-dimensional  Rashba systems\\ 
with in-plane Zeeman field and point-like disorder}

\author{Igor Gornyi}
 \affiliation{\mbox{Institute for Quantum Materials and Technologies, Karlsruhe Institute of Technology, 76131 Karlsruhe, Germany}} 
 \affiliation{\mbox{Institut f\"ur Theorie der Kondensierten Materie, Karlsruhe Institute of Technology, 76131 Karlsruhe, Germany}}

 \author{Alexander Khaetskii}
 \affiliation{Department of Physics and Astronomy, Ohio University, Athens, OH, USA}

 \date{\today}

\begin{abstract}
We study the dc conductivity of a two-dimensional Rashba system
with an in-plane Zeeman field and delta-correlated scalar disorder.
We show that, although the field deforms the two helicity Fermi contours
and rotates the spin texture, it does not produce anisotropic 
magnetoresistance in the leading quasiclassical conductivity. 
The result follows from a geometric identity for the total area occupied by the two helicity sheets. A density Ward identity fixes the
spin-vector part of the Born self-energy to the derivative of the total
particle density with respect to the field. This derivative vanishes,
because the total area enclosed by the two Rashba-Zeeman sheets is
independent of the in-plane field. The Born self-energy is therefore
scalar and field independent, and the quasiparticle lifetime stays
isotropic. The same occupied-area identity controls transport: the leading
impurity ladder reduces the current vertex to the parabolic velocity, 
and the diagonal intraband Kubo conductivity collapses onto the 
two-sheet occupied area and is field independent as well. 
Thus, point-like nonmagnetic impurities produce neither angular
anisotropy nor field-magnitude dependence in the leading diagonal
two-sheet quasiclassical conductivity. Any residual AMR from
delta-correlated disorder must originate from corrections outside this sector.

\end{abstract}

\maketitle

\section{Introduction}
\label{sec:intro}

Anisotropic magnetoresistance (AMR)---the dependence of electrical
resistance on the direction of an applied magnetic field or
magnetization---is a long-standing phenomenon~\cite{Thomson1857} with 
both fundamental and technological importance. In low-dimensional
conductors with spin--orbit coupling, such as semiconductor and oxide
two-dimensional electron gases, an in-plane Zeeman field breaks
spin-rotation symmetry, changes the momentum-dependent spin orientation
of the electronic states, and deforms the helicity Fermi contours. This
makes Rashba systems with an in-plane field a natural minimal setting
for discussing AMR without magnetic order. The central question is
which microscopic ingredients are sufficient to produce a finite AMR.

Previous studies have reached different conclusions from different angles. 
Raimondi \emph{et al.}~\cite{Raimondi2001} studied essentially the
same parabolic two-dimensional Rashba--Zeeman model with scalar
short-range disorder and found that impurity vertex corrections
strongly modify the AMR, including its sign. Their work therefore
makes this model a particularly useful setting for asking what
remains of the anisotropy in the leading weak-disorder limit.

Inoue \emph{et al.}~\cite{Inoue2006} showed, for the anomalous Hall response
of a spin-polarized Rashba model, that the intrinsic contribution is
removed by vertex corrections unless the quasiparticle lifetime is
spin-dependent. Kato \emph{et al.}~\cite{Kato2008} formulated the
analogous cancellation criterion for intrinsic AMR. Trushin
\emph{et al.}~\cite{Trushin2009} studied carriers with Rashba and
Dresselhaus couplings scattered by polarized magnetic impurities,
thus addressing a different magnetic-disorder mechanism. Wang and
Pang~\cite{WangPang2009} performed a numerical study that made a sharp
distinction between point-like and remote nonmagnetic impurities: the
AMR vanished for delta-function disorder but became finite for
long-range disorder. Wang later extended this analysis to combined
Rashba--Dresselhaus systems and again emphasized that, for short-range
spin-independent disorder, AMR vanishes when both helicity bands are
occupied, while this cancellation no longer applies in the single-band
regime~\cite{WangPang2010}. The long-range and nonparabolic mechanisms
were explored further in Ref.~\cite{WangPang2011}. Vaz
\emph{et al.}~\cite{Vaz2020} used bilinear magnetoresistance to extract
an effective Rashba parameter in an oxide two-dimensional electron gas.
Their microscopic theory is directly relevant here because its
minimal Rashba model, without additional orbital bands, also uses a 
local scalar impurity potential, a Born self-energy, and impurity 
vertex corrections, but it contains a field-dependent relaxation-time 
structure. 
Soori~\cite{Soori2021} considered finite transverse conductance and anisotropic magnetoconductance in 2DEGs with strong spin--orbit coupling, stressing the importance of Fermi-surface geometry.

Together, these results leave a concrete puzzle. For point-like
disorder, some calculations find no AMR, whereas others obtain a finite 
anisotropy or a field-dependent scalar lifetime. 
Furthermore, the numerical works that find zero AMR for delta-correlated 
disorder mainly address angular anisotropy, not the stronger question whether
the diagonal conductivity itself is independent of the
field magnitude. Both issues arise in models with the same Rashba--Zeeman Hamiltonian, point-like disorder, and impurity ladder. 
The present paper addresses both issues for the leading diagonal two-sheet
quasiclassical sector. In this sector, the scalar Born lifetime is not
an independent angle-dependent input, and the leading diagonal
conductivity is not a generic functional of the separately distorted
contours. Both quantities are fixed by the same geometric identity. As
a result, the quasiclassical conductivity does not depend on the
in-plane magnetic field.

The first part of the argument concerns the Born self-energy. For
short-range scalar disorder, the self-energy is the full
momentum-integrated on-shell spectral matrix. Its spin-vector part is
related to the derivative of the total density with respect to the
in-plane Zeeman field. This identity is the Zeeman analogue of the
thermodynamic logic behind St\v{r}eda's formula~\cite{Streda1982}: a
Fermi-sea derivative fixes a Fermi-surface response. In the 
parabolic Rashba model with both helicity sheets included, the total
occupied area is unchanged by the displacement of the Rashba degeneracy
point. Therefore, the spin-vector part of the Born broadening vanishes
and the projected quasiparticle lifetime is the field-independent
scalar lifetime $\tau_0$. The perturbative calculation in Sec.~S2 of
the Supplemental Material shows explicitly how an apparent anisotropic
lifetime is removed once the density-of-states factor and the spin
projector are evaluated on the same helicity Fermi sheet.

The second part of the argument concerns the current vertex. A
point-like impurity potential does not mean that the transport vertex is
bare in a spin--orbit-coupled band. The spinor overlap makes the
projected transport kernel nontrivial. The impurity ladder cancels the 
anomalous Rashba part of the bare velocity and leaves the parabolic current
vertex. This cancellation follows from derivatives of the same occupied-area
identity that controls the Born self-energy. With this dressed vertex, the
diagonal intraband conductivity reduces to the total area occupied by the 
two helicity sheets and is therefore independent of the field. Thus, the
Fermi-contour deformation, the spin-texture rotation, the scalar Born
lifetime, and the impurity ladder together give no AMR for point-like
scalar disorder; in fact, within this diagonal quasiclassical sector,
they give no field dependence of the longitudinal conductivity.

The paper is organized as follows. Section~\ref{system} introduces the
Rashba--Zeeman Hamiltonian and the two-sheet regime.
Section~\ref{sec:scattime} derives the Born self-energy and the Ward
identity for the lifetime. Section~\ref{Sec:Conductivity}
evaluates the diagonal intraband Kubo conductivity and reduces it to
the total occupied area. Section~\ref{sec:comparison} compares the
result with previous work and identifies which short-range-disorder
claims are consistent with the present Ward-identity calculation.
Appendix~\ref{app:dos_independence} proves the area identity, and
Appendix~\ref{sec:vertex} gives the impurity-ladder vertex. The
Supplemental Material (SM)~\cite{SM} gives the projected Dyson
derivation, the perturbative weak-field check of the lifetime
cancellation, and an alternative shifted-coordinate proof of the area
identity.


\section{The system}
\label{system}
\subsection{Hamiltonian}
\label{Sec:Hamiltonian}

The Hamiltonian of the model and the velocity operator are:
\begin{align}
\hat{H}(\vb{p}) = \frac{\vb{p}^{2}}{2m}\hat{I}
 + \alpha\hat{\vb{\sigma}}\cdot\vb{p}
 + \vb{B}\cdot\hat{\vb{\sigma}};
 \label{eq:Ham}
\\
\hat{\vb{v}}(\vb{p})
 = \frac{\partial\hat{H}}{\partial\vb{p}}
 = \frac{\vb{p}}{m}\hat{I} + \alpha\hat{\vb{\sigma}}.
 \label{eq:vb}
\end{align}
Here, $\vb p$ is the two-dimensional electron momentum, $\alpha$ is
the Rashba spin-orbit coupling, and $\hat{\vb\sigma}$ are the Pauli
matrices. Bold symbols denote two-dimensional vectors. 
The orbital
coupling to the magnetic field is absent; only the in-plane Zeeman term
is retained.
We write the Zeeman coupling entering the Hamiltonian as an
energy-valued vector
\begin{equation}
\vb B = B\,\vb e_B,
\quad
\vb e_B=(\cos\phi_B,\sin\phi_B),
\quad
B=|\vb B|.
\label{B}
\end{equation}
The sign of the microscopic Zeeman coupling has been absorbed into
$\vb B$. If $\mathcal{B}$ denotes the magnitude of the physical
magnetic field, then $B=g\mu_B \mathcal{B}/2$ up to this sign
convention. 

The form $\alpha\hat{\vb\sigma}\cdot\vb p$ is obtained from the
conventional Rashba coupling
$\alpha(\hat{\vb\sigma}\times\vb p)_z
=\alpha(\hat\sigma_xp_y-\hat\sigma_yp_x)$
by a fixed in-plane rotation of the spin basis. In particular, defining
$\tilde\sigma_x=-\hat\sigma_y$ and
$\tilde\sigma_y=\hat\sigma_x$ gives
$\alpha(\hat{\vb\sigma}\times\vb p)_z
=\alpha\tilde{\vb\sigma}\cdot\vb p$.
The Zeeman vector $\vb B$ in Eq.~\eqref{eq:Ham} is expressed in the
same rotated spin basis.

In what follows, we use units with $\hbar=1$.

The eigenvalues of the Hamiltonian are
\begin{align}
\epsilon_{\pm}(\vb{p})
&= \frac{p^{2}}{2m}
 \pm|\alpha\vb{p}+\vb{B}| \nonumber\\
&= \frac{p^{2}}{2m}
 \pm\sqrt{(\alpha p)^{2}+B^{2}+2\alpha pB\cos\theta},
\label{eq:eps}
\end{align}
where $\theta$ is the angle between $\vb{p}$ and $\vb{B}$.
The band velocities obtained from the dispersion are
\begin{align}
\vb v_{\pm}
&=
\frac{\vb p}{m}\pm\alpha\vb\nu_{\vb p}, \quad 
\vb\nu_{\vb p}(\vb B) =
\frac{\alpha\vb p+\vb B}
{|\alpha\vb p+\vb B|},
\label{nup}
\end{align}
where $\vb\nu_{\vb p}$ is the unit vector in the direction of the
local Rashba-Zeeman field. The two branches of the energy spectrum
correspond to positive and negative eigenvalues of the chirality
operator
\begin{equation}
\hat\nu(\vb p)
=
\vb\nu_{\vb p}\cdot\hat{\vb\sigma}.
\label{eq:chiral}
\end{equation}
The projectors onto these branches are
\begin{equation}
\hat\Omega^{\pm}(\vb p)
=
\frac{1\pm\hat\nu(\vb p)}{2}.
\label{eq:Omega}
\end{equation}

We consider a short-ranged impurity potential
\[V_{\text{imp}}=\sum_{i}U\delta(\vb{r}-\vb{R}_{i}).\]
In the absence of Zeeman field the Born scattering rate is 
\begin{equation}\frac1{\tau_0}
=
2\pi n_{\rm imp}|U|^2\nu_0
=
{n_{\rm imp}|U|^2m},
\label{eq:tau0def}
\end{equation}
where $n_{\text{imp}}$ is the 2D impurity concentration
and $
\nu_0=m/2\pi
$
is the density of states per spin in two dimensions.

We distinguish the actual chemical potential $\mu$ from the reference Fermi energy $E_F$. The former is measured from the
Rashba degeneracy energy at $p=0$, for which
$\epsilon_\pm(0)=0$ at $B=0$, whereas $E_F$ is the Fermi energy of the
spin-degenerate parabolic band at $\alpha=0$ at the same carrier
concentration:
\begin{equation}
 E_F=\frac{\pi n}{m}=\frac{p_F^2}{2m},
 \qquad
 p_F=\sqrt{2mE_F}.
\label{eq:EF_reference}
\end{equation}
In the present two-sheet Rashba model, 
\begin{equation}
 E_F=\mu+m\alpha^2.
\label{eq:mu_EF_relation}
\end{equation}
Throughout the paper, we work in the clean metallic regime
$E_F\tau_0\gg1$ and $\alpha p_F\tau_0\gg1$. The additional
conditions that both helicity sheets are present and that the
diagonal-helicity projection is controlled are specified in
Sec.~\ref{sec:spectrum_lifshitz}.

\subsection{Spectrum}
\label{sec:spectrum_lifshitz}

\begin{figure}[!ht]
\hspace{0.3cm}
\includegraphics[width=0.9\columnwidth]{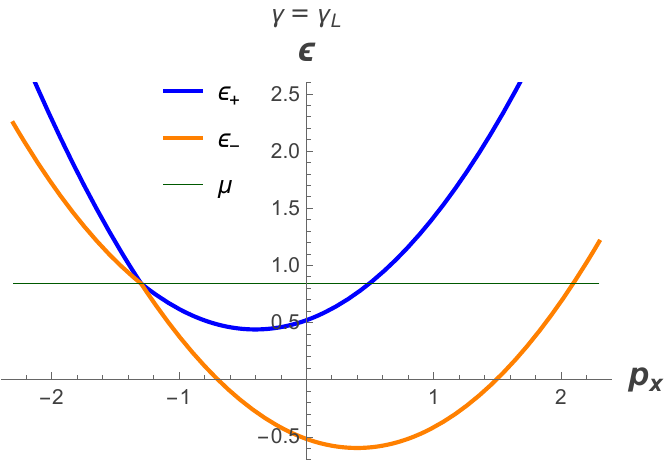}
\vspace{0.5cm}
\\
\hspace{-1cm}
\includegraphics[width=0.75\columnwidth]{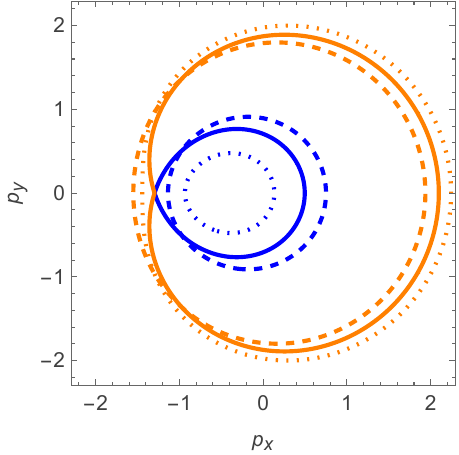}
\caption{\label{Fig:Spectrum}
Spectrum and Fermi contours for
$\alpha=0.4\sqrt{E_F/m}$, for which
$\gamma_L\simeq0.917$. Energies are measured in units of $E_F$ and momenta in units of
$\sqrt{mE_F}$.
Upper panel: helicity dispersions along the field axis
($p_y=0$, $\vb B\parallel\vb e_x$) at
$\gamma=\gamma_L$. The two branches
$\epsilon_+$ (blue) and $\epsilon_-$ (orange) meet at the
Rashba--Zeeman degeneracy point
$p_x=p_*=-B/\alpha$, which lies at the Fermi level $\mu$
at the Lifshitz crossing.
Lower panel: Fermi contours for
$\gamma=\gamma_L/2$ (dashed),
$\gamma=\gamma_L$ (solid), and
$\gamma=3\gamma_L/2$ (dotted), with the same colors for the
two helicity branches. The individual contours and their enclosed
areas change with the field, while the sum of the two occupied areas
remains field independent throughout the two-sheet regime (Appendix~\ref{app:dos_independence}).
}
\end{figure}

The local band splitting is $$
\Delta(\vb p)=2|\alpha\vb p+\vb B|.
$$
It vanishes at the Rashba-Zeeman degeneracy point
\begin{equation}
\vb p_*=-\frac{\vb B}{\alpha}.
\label{eq:pstar_gamma_one}
\end{equation}
This point exists for every value of the in-plane field. 
At zero field, the Fermi level reaches the Rashba degeneracy
(Dirac) point at $\mu=0$, which by
Eq.~\eqref{eq:mu_EF_relation} corresponds to $E_F=m\alpha^2$.
What changes
with the field is whether the degeneracy point lies on the Fermi
surface. Its energy is purely kinetic,
\begin{equation}
E_*=
\frac{p_*^2}{2m}
=
\frac{B^2}{2m\alpha^2}.
\end{equation}
Using $p_F^2=2mE_F$, we obtain

\begin{equation}
E_* = \gamma^2 E_F,
\quad
\gamma=\frac{B}{\alpha p_F}.
\label{eq:Estar_gamma_one}
\end{equation}
The Fermi surface passes through $\vb p_*$ when $E_*=\mu$, i.e.,
at
\begin{equation}
 \gamma=\gamma_{\rm L},
 \quad
 \gamma_{\rm L}
 =
 \sqrt{\frac{\mu}{E_F}}
 =
 \sqrt{1-\frac{m\alpha^2}{E_F}}.
\label{eq:gamma_L}
\end{equation}
Thus, $E_*<\mu$ for $\gamma<\gamma_{\rm L}$,
$E_*=\mu$ for $\gamma=\gamma_{\rm L}$, and
$E_*>\mu$ for $\gamma>\gamma_{\rm L}$.
The equality $\gamma=\gamma_{\rm L}$ defines the Lifshitz crossing
relevant for the on-shell quasiparticle problem. For larger fields,
$\vb p_*$ remains the Rashba--Zeeman degeneracy point but lies off
the Fermi surface. The upper panel of Fig.~\ref{Fig:Spectrum} shows the Lifshitz
crossing: at $\gamma=\gamma_{\rm L}$ the two helicity branches meet
at the Fermi level at $\vb p=\vb p_*$.

Near the antiparallel direction, we write
$
\theta=\phi-\phi_B=\pi+\eta,
\,\,\,
|\eta|\ll1.
$

At the crossing, where the degeneracy momentum is
$\sqrt{2m\mu}=p_F\gamma_L$, the local splitting has the form
\begin{equation}
\Delta(\eta)
\simeq
2\alpha p_F
\sqrt{(\gamma-\gamma_L)^2
+\gamma\gamma_L\eta^2}.
\label{eq:Delta_eta_gamma_one}
\end{equation}
In particular,
\begin{equation}
\Delta_{\min}
\simeq
2\alpha p_F|\gamma-\gamma_L|,
\end{equation}
and it vanishes at $\gamma=\gamma_L$.
The lower panel of Fig.~\ref{Fig:Spectrum} shows how the two helicity
Fermi contours evolve across this crossing. Although their shapes and
individual enclosed areas change with the field, their total occupied
area remains unchanged in the two-sheet regime, as proved in
Appendix~\ref{app:dos_independence}.

We will work in the two-sheet regime throughout the main calculation.
By this we mean that both helicity branches cross the chemical
potential for every direction in momentum space, so that each branch is
described by a single positive outer Fermi radius $
\vb p_s(\phi)=p_s(\phi)\vb e_\phi,
\,\, s=\pm .$

In particular, the $s=+$ sheet is present only when
\[
 B<\mu=E_F-m\alpha^2 .
\]

At $\gamma=\gamma_L$, the Rashba--Zeeman degeneracy lies on
the Fermi surface and the local helicity splitting vanishes. The
diagonal-helicity projection is controlled only when
$\Delta_{\min}\tau_0\gg1$. Using the expression for
$\Delta_{\min}$ above, this excludes the disorder-broadened
neighborhood
\begin{equation}
|\gamma-\gamma_L|\lesssim\epsilon_\tau,
\qquad
\epsilon_\tau=\frac{1}{2\alpha p_F\tau_0}.
\label{eq:lifshitz_window}
\end{equation}
The calculation is therefore restricted to
$B<\mu$ and $|\gamma-\gamma_L|\gg\epsilon_\tau$.
The second condition is relevant only where the excluded
disorder-broadened interval overlaps the two-sheet domain.

\section{Dyson equation and scalar Born lifetime}
\label{sec:scattime}

We now derive the scalar lifetime that enters the diagonal conductivity
calculation. The important point is that the short-range Born
self-energy is momentum independent but remains a matrix in spin space.
When this matrix is projected onto the local helicity eigenstates, an
angle-dependent lifetime is possible in a generic spin-orbit model. In
the present Rashba model, however, the spin-vector part of the
on-shell Born self-energy vanishes by a Ward identity. The resulting
scalar lifetime is isotropic.

\subsection{Born self-energy and projected scattering rate}
\label{sec:born_projected_rate}

For a short-range scalar impurity potential,
$
\langle U(\vb r)U(\vb r')\rangle
=
n_{\rm imp}|U|^2\delta(\vb r-\vb r'),
$
the Born self-energy is
\begin{equation}
\hat\Sigma^R(\epsilon)
=
n_{\rm imp}|U|^2
\int\frac{d^2p}{(2\pi)^2}\,
\hat G^{R,0}(\epsilon,\vb p).
\label{eq:Sigma_def_main}
\end{equation}
Here the clean retarded Green's function is
\begin{equation}
\hat G^{R,0}(\epsilon,\vb p)
=
\frac{\hat\Omega_+(\vb p)}
 {\epsilon-\epsilon_+(\vb p)+i0^+}
+
\frac{\hat\Omega_-(\vb p)}
 {\epsilon-\epsilon_-(\vb p)+i0^+}.
\label{eq:G0_app}
\end{equation}
Evaluating the radial
integral on shell at $\epsilon=\mu$, one obtains
\begin{equation}
-\operatorname{Im}\hat\Sigma^R(\mu)
=
\pi n_{\rm imp}|U|^2
\sum_{s=\pm}
\int\frac{d^2p}{(2\pi)^2}
\delta(\mu-\epsilon_s(\vb p))\,
\hat\Omega_s(\vb p),
\label{eq:ImSigma_spectral_main}
\end{equation}
where the projector operator is given by Eq. (\ref{eq:Omega}), and 
on the Fermi contour of branch $s$ we write (with $\vb p_s(\phi)=p_s(\phi)\vb e_\phi$, $\vb e_\phi=(\cos \phi,\sin\phi)$, and $\vb e_\theta=(-\sin\phi,\cos\phi)$)
\begin{equation}
\vb\nu^{(s)}(\phi)
\equiv
\vb\nu_{\vb p}\big|_{\vb p=\vb p_s(\phi)}
=
\frac{\alpha p_s(\phi)\vb e_\phi+\vb B}
{\left|\alpha p_s(\phi)\vb e_\phi+\vb B\right|}.
\label{eq:nu_s_on_shell}
\end{equation}
Projecting Eq.~\eqref{eq:ImSigma_spectral_main} onto a local helicity
state $|s,\phi\rangle$ gives the branch-resolved Golden-Rule rate (see Sec. S1 of SM for details)
\begin{align}
\frac{1}{\tau_s(\phi)}
&=
2\pi n_{\rm imp}|U|^2\notag \\
&\times \sum_{s'=\pm}
\int_0^{2\pi}\frac{d\phi'}{2\pi}\,
\rho_{s'}(\phi')
\frac{
1+ss'\,\vb\nu^{(s)}(\phi)\cdot\vb\nu^{(s')}(\phi')
}{2}.
\label{eq:tau_def_corrected_main}
\end{align}
Here,
\begin{equation}
\rho_s(\phi)
=
\frac{p_s(\phi)}
{2\pi |v_{s,\parallel}(\phi)|},
\quad
v_{s,\parallel}(\phi)
=
\left.
\frac{\partial\epsilon_s(p,\phi)}{\partial p}
\right|_{p=p_s(\phi)} 
\label{eq:rho_main}
\end{equation}
are the angular density of states and the radial velocity on sheet $s$. The spin texture must be evaluated on the same sheet. The correction to the spin texture
caused by the branch-dependent displacement of the Fermi sheet is of
the same order as the correction to the angular density of states and cancels the apparent anisotropy. 
The detailed projection algebra is given in Sec. S1 of the
SM. 

\subsection{Ward identity behind the isotropic lifetime}
\label{sec:tau_ward_identity}

The cancellation of the anisotropic scalar lifetime is not an accident
of the low-order expansion. It follows from a Ward identity that
relates the spin-vector part of the Born broadening to the derivative
of the total particle density with respect to the in-plane Zeeman field.

We decompose the on-shell Born broadening as
\begin{equation}
-\operatorname{Im}\hat\Sigma^R
=
\Gamma_0\hat{\mathds{I}}+\vb\Gamma\cdot\hat{\vb\sigma}.
\label{eq:Gamma_decomp_main}
\end{equation}
The scalar part is proportional to the total on-shell density of states,
while the spin-vector part is
\begin{equation}
\vb\Gamma
=
\frac{\pi n_{\rm imp}|U|^2}{2}
\sum_s s
\int\frac{d^2p}{(2\pi)^2}
\delta(\mu-\epsilon_s)
\vb\nu_{\vb p}.
\label{eq:Gamma_vector_main}
\end{equation}
A quasiparticle on branch $s$ and angle
$\phi$ sees the projection of this total matrix,
\begin{equation}
\frac{1}{2\tau_s(\phi)}
=
\Gamma_0+s\,\vb\Gamma\cdot\vb\nu^{(s)}(\phi).
\label{eq:band_rate_from_total_matrix_main}
\end{equation}

For the Rashba-Zeeman spectrum
$$
\epsilon_s(\vb p)
=
\frac{p^2}{2m}
+
s|\alpha\vb p+\vb B|,
$$
one has
$
\partial\epsilon_s/\partial\vb B
= s\vb\nu_{\vb p}$.
Therefore the spin-vector spectral weight in
Eq.~\eqref{eq:Gamma_vector_main} can be written as
\begin{align}
\vb M_{\rm FS}(\mu,\vb B)
&\equiv 
\sum_s s
\int\frac{d^2p}{(2\pi)^2}
\delta(\mu-\epsilon_s)
\vb\nu_{\vb p} \nonumber \\
&=
\sum_s
\int\frac{d^2p}{(2\pi)^2}
\delta(\mu-\epsilon_s)
\frac{\partial\epsilon_s}{\partial\vb B}.
\label{eq:MFS_def_main}
\end{align}
On the other hand, the total density is
\begin{equation}
n(\mu,\vb B)
=
\sum_s
\int\frac{d^2p}{(2\pi)^2}
\Theta(\mu-\epsilon_s(\vb p)).
\label{eq:n_total_def_main}
\end{equation}
Differentiating it with respect to $\vb B$ gives
\begin{equation}
\vb M_{\rm FS}(\mu,\vb B)
=
-\frac{\partial n(\mu,\vb B)}{\partial\vb B}.
\label{eq:M_density_Ward}
\end{equation}

Equation~\eqref{eq:M_density_Ward} is the Ward identity associated
with the continuous momentum-space displacement symmetry of the
two-sheet Fermi-sea measure established in
Appendix~\ref{app:dos_independence}. This symmetry plays the same
structural role as gauge symmetry in the Ward--Takahashi
identities: its infinitesimal action relates a Fermi-sea derivative
to the corresponding on-shell vertex insertion. Here, varying the
in-plane Zeeman field displaces the Rashba degeneracy in momentum
space.
It can also be viewed as the Zeeman analogue of the thermodynamic 
logic behind St\v{r}eda's formula~\cite{Streda1982}: a derivative of 
an equilibrium Fermi-sea quantity fixes a Fermi-surface spectral 
response.   
In the parabolic model with both helicity
sheets included, the total occupied area is independent of the
displacement of the Rashba degeneracy point by the in-plane field,
$
\partial n(\mu,\vb B)/\partial\vb B=0.
$
The proof is given in Appendix~\ref{app:dos_independence}. 
With the standard continuum
convention, Eqs.~(\ref{eq:Atot_final}) and (\ref{density}) give

\begin{equation}
n(\mu,\vb B)=n(\mu,0)
=
\frac{m\mu}{\pi}
+
\frac{m^2\alpha^2}{\pi}
=
\frac{mE_F}{\pi}
=
\frac{p_F^2}{2\pi}.
\label{eq:n_independent_of_B}
\end{equation}
Since $\partial n/\partial\vb B=0$ and
$\partial n/\partial\mu=m/\pi>0$, fixing the carrier concentration
also fixes $\mu$ independently of $\vb B$. The conclusions below are
therefore identical at fixed chemical potential and at fixed density.

Thus, the spin-vector part of the on-shell Born
broadening vanishes:
\begin{equation}
\vb\Gamma=0,
\qquad
-\operatorname{Im}\hat\Sigma^R=\Gamma_0\hat{\mathds{I}}.
\label{eq:Gamma_vector_zero_main}
\end{equation}
The branch-projected scalar lifetime is therefore isotropic and field
independent:
\begin{equation}
\frac{1}{\tau_s(\phi)}
=
2\Gamma_0
\equiv
\frac{1}{\tau_0}.
\label{eq:tau_isotropic_Ward_result}
\end{equation}
The scalar lifetime result is exact within the two-sheet, on-shell Born model for short-range disorder.

\medskip

\section{Conductivity}
\label{Sec:Conductivity}

\subsection{General framework}

We start from the retarded--advanced part of the disorder-averaged
Kubo formula,
\begin{equation}
\sigma_{ij}
 =
 \left\langle
 \frac{e^{2}}{2\pi}
 \int\frac{d^{2}\vb{p}}{(2\pi)^{2}}\,
 \operatorname{tr}\bigl[
 \hat{v}_{i}(\vb{p})\,G^{A}(\mu,\vb{p})\,
 \hat{v}_{j}(\vb{p})\,G^{R}(\mu,\vb{p})
 \bigr]
 \right\rangle .
\label{eq:Kubo}
\end{equation}

Here, the trace is over spin and $\langle\ldots\rangle$ denotes disorder averaging. In the regime
$E_F\tau_0\gg1$, the retarded--advanced on-shell term gives the leading
metallic dc conductivity. Retarded--retarded, advanced--advanced, and
Fermi-sea terms lack this enhancement and are smaller by a relative
factor of order $(E_F\tau_0)^{-1}$. The diagonal-helicity projection
used below is controlled by $\Delta_{\min}\tau_0\gg1$; corrections
involving mixing of the two helicity branches are suppressed by
inverse powers of $\Delta_{\min}\tau_0$. The projection therefore
ceases to be uniform in the Lifshitz window excluded in Eq.~\eqref{eq:lifshitz_window}.

With the scalar broadening inserted, the diagonal helicity Green's
functions used in this section are
\begin{equation}
\hat G^{R,A}(\mu,\vb p)
=
\sum_{s=\pm}
\frac{\hat\Omega_s(\vb p)}
 {\mu-\epsilon_s(\vb p)\pm i/(2\tau_0)}.
\label{eq:GRA}
\end{equation}
Using the leading ladder correction derived in
Appendix~\ref{sec:vertex}, we write
$
\hat V_i^{\rm lead}=p_i/m
$
for the dressed vertex in the diagonal intraband calculation below.
Keeping only intraband Kubo
pairs, we obtain
\begin{widetext}
\begin{align}
\sigma_{ij}
 =
 \frac{e^{2}}{2\pi}
 \int\frac{d^{2}\vb{p}}{(2\pi)^{2}}\,
 \left(\frac{\vb p}{m}\right)_i
 \left[
 \frac{\left({\vb p}/{m}+\alpha\vb\nu_{\vb p}\right)_j}
 {[\mu-\epsilon_+(\vb p)-i/2\tau_0]
 [\mu-\epsilon_+(\vb p)+i/2\tau_0]}
 +
 \frac{\left({\vb p}/{m}-\alpha\vb\nu_{\vb p}\right)_j}
 {[\mu-\epsilon_-(\vb p)-i/2\tau_0]
 [\mu-\epsilon_-(\vb p)+i/2\tau_0]}
 \right].
\label{eq:sigmaij}
\end{align}
Here, $\vb\nu_{\vb p}$ is defined by Eq.~\eqref{nup}. 

Integrating over $\xi=\epsilon_s(\vb p)-\mu$ on each helicity branch,
we obtain the Fermi-surface contribution
\begin{equation}
\sigma_{ij}
 =
 e^{2}\tau_0
 \int_{0}^{2\pi}\frac{d\phi}{(2\pi)^{2}}\,
 \left\{
 \frac{p\,(\vb p/m)_i\,(\vb p/m+\alpha\vb\nu_{\vb p})_j}
 {p/m+\alpha\vb\nu_{\vb p}\cdot\vb e_{\phi}}
 \bigg|_{p=p_+}
 +
 \frac{p\,(\vb p/m)_i\,(\vb p/m-\alpha\vb\nu_{\vb p})_j}
 {p/m-\alpha\vb\nu_{\vb p}\cdot\vb e_{\phi}}
 \bigg|_{p=p_-}
 \right\}.
\label{eq:sigmaFS}
\end{equation}
Here, $p_\pm(\phi)$ are the solutions of
$\epsilon_\pm(p,\phi)=\mu$. All quantities in each term, including
$\vb\nu_{\vb p}$, are evaluated at the same on-shell momentum $p=p_\pm(\phi)$.

\subsection{Diagonal quasiclassical conductivity}
\label{sec:arbitrary}

We now evaluate the diagonal intraband contribution to the
quasiclassical dc conductivity in the split-band regime where the
helicity projection is controlled. The band velocity on helicity
branch $s$ is
\begin{equation}
\vb v_s(\vb p)=\nabla_{\vb p}\epsilon_s(\vb p)
=
\frac{\vb p}{m}
+
s\alpha\vb\nu_{\vb p}.
\label{eq:band_velocity_vs}
\end{equation}
Its radial component $v_{s,\parallel}=\vb v_s\cdot\vb e_\phi$ is
\begin{align}
v_{s,\parallel}(\phi)
&=
\left.
\frac{\partial\epsilon_s}{\partial p}
\right|_{p=p_s(\phi)}
=
\frac{p_s(\phi)}{m}
+
s\alpha\,\vb\nu^{(s)}(\phi)\cdot\vb e_\phi, \qquad \vb\nu^{(s)}(\phi)
=
\vb\nu_{\vb p}\big|_{\vb p=p_s(\phi)\vb e_\phi}.
\label{eq:vs_parallel_geometric}
\end{align}
Thus, the denominator in the $s$-branch term of
Eq.~\eqref{eq:sigmaFS} is precisely the radial velocity
$v_{s,\parallel}(\phi)$. Then the $s$-branch contribution to Eq.~\eqref{eq:sigmaFS} can be
rewritten as
\begin{align}
\sigma_{xx}^{(s)}
=
e^2\tau_0
\int_0^{2\pi}\frac{d\phi}{(2\pi)^2}
p_s(\phi)
\frac{
(\vb p/m)_x\,(\vb p/m+s\alpha\vb\nu_{\vb p})_x
}
{p/m+s\alpha\,\vb\nu_{\vb p}\cdot\vb e_\phi}
\bigg|_{\vb p=p_s(\phi)\vb e_\phi}
=
e^2\tau_0
\int_0^{2\pi}\frac{d\phi}{(2\pi)^2}\,
p_s(\phi)\,
\frac{
\left[p_s(\phi)\cos\phi/m\right]\,v_{s,x}(\phi)
}
{v_{s,\parallel}(\phi)} .
\label{eq:sigma_branch_contour}
\end{align}
\end{widetext}
The Fermi contour is defined by $
\epsilon_s\left[p_s(\phi),\phi\right]=\mu . $
Differentiating this identity with respect to $\phi$ gives
\begin{align}
\frac{d}{d\phi}
\epsilon_s\left[p_s(\phi),\phi\right]
=
\left.
\frac{\partial \epsilon_s}{\partial \phi}
\right|_{p=p_s(\phi)}
+
v_{s,\parallel}
\frac{d p_s}{d\phi} = 0.
\label{eq:contour_derivative}
\end{align}
On the Fermi contour, the azimuthal velocity is
\begin{equation}
v_{s,\phi}(\phi)
=
\vb v_s\cdot\vb e_\theta
=
\frac{1}{p_s(\phi)}
\left.
\frac{\partial \epsilon_s}{\partial \phi}
\right|_{p=p_s(\phi)}
= -
\frac{v_{s,\parallel}(\phi)}{p_s(\phi)} \frac{d p_s(\phi)}{d\phi},
\label{eq:vphi_contour}
\end{equation}
and 
the $x$-component of the velocity is
\begin{align}
v_{s,x}(\phi)
&=
v_{s,\parallel}(\phi)\,\cos\phi
-
v_{s,\phi}(\phi)\,\sin\phi 
\notag
\\
&=
v_{s,\parallel}(\phi)
\left[
\cos\phi+
\frac{\sin\phi}{p_s(\phi)}\,\frac{d p_s(\phi)}{d\phi}
\right].
\label{eq:vx_over_vparallel}
\end{align}
Using Eq.~\eqref{eq:vx_over_vparallel} in
Eq.~\eqref{eq:sigma_branch_contour}, we obtain

\begin{align}
\sigma_{xx}^{(s)}\!=\!
\frac{e^2\tau_0}{m}\!
\int_0^{2\pi}\!\!\!\frac{d\phi}{(2\pi)^2}\,p_s(\phi)\cos\phi\!
\left[
p_s\cos\phi 
 + 
\frac{d p_s(\phi)}{d\phi}
\sin\phi
\right]\!.
\label{eq:sigma_branch_expanded}
\end{align}
The second term in the brackets appears because the band velocity is normal to the
distorted Fermi contour rather than purely radial.
This term produces an area contribution to $\sigma_{xx}^{(s)}$ after integration by parts:

\begin{align}
\int_0^{2\pi}\!d\phi\,
p_s(\phi)\frac{d p_s(\phi)}{d\phi}\,\sin\phi\cos\phi
\!=\!
-\frac12
\int_0^{2\pi}\!d\phi\,
p_s^2(\phi)\cos2\phi .
\label{eq:integration_by_parts_contour}
\end{align}
 Combining
this with Eq.~\eqref{eq:sigma_branch_expanded} gives
\begin{align}
\sigma_{xx}^{(s)}
=
\frac{e^2\tau_0}{2m}
\int_0^{2\pi}\frac{d\phi}{(2\pi)^2}\,
p_s^2(\phi).
\label{eq:sigma_branch_area_form}
\end{align}

Thus, each branch contributes only through the area enclosed by its
Fermi contour. Summing over the two helicity sheets gives
\begin{equation}
\sigma_{xx}
\!=\!
\frac{e^2\tau_0}{(2\pi)^2}
\frac{1}{2m}
\int_0^{2\pi}\!d\phi\,
\left[p_+^2(\phi)+p_-^2(\phi)\right]
\!=\!
\frac{e^2\tau_0}{(2\pi)^2m}
\,A_{\rm tot}^{(p)}.
\label{eq:sigma_area_form}
\end{equation}
Hence, within the scalar-lifetime and leading-vertex diagonal
approximation, $\sigma_{xx}$ is determined only by the total occupied
area of the two helicity sheets.
Using the density-area relation in Eq.~\eqref{density},
Eq.~\eqref{eq:sigma_area_form} takes the Drude form 
\begin{equation}
\sigma_{xx}(B)
=
\frac{e^2n(\mu,\vb B)\tau_0}{m}.
\label{eq:sigma_drude_area}
\end{equation}

At $B=0$, this result agrees with
Refs.~\cite{Brosco2016,Golub2016}.

The area identity proved in Appendix~\ref{app:dos_independence} gives
$n(\mu,\vb B)=n(\mu,0)$, and therefore 

\begin{equation}
\sigma_{xx}(B)
=
\sigma_{xx}(0)
=
\frac{e^2p_F^2\tau_0}{2\pi m}
=
\frac{e^2E_F\tau_0}{\pi}
=
\frac{e^2n\tau_0}{m}.
\label{eq:sig_arb}
\end{equation}

Thus, $\sigma_{xx}$ is independent of $\vb B$.
The same contour argument, with $x$ replaced by $y$, gives
$
\sigma_{yy}(B)=\sigma_{yy}(0)=\sigma_{xx}(0),
$
and hence 
\begin{equation}
\sigma_{xx}-\sigma_{yy}=0.
\label{main-result}
\end{equation}
Equations~\eqref{eq:sig_arb} and~\eqref{main-result} are the central
transport result of the paper.

The conventional AMR is stated in terms of the resistivity. Choosing
the axes parallel and perpendicular to the in-plane field,
Eq.~\eqref{main-result} reads
$
\sigma_\parallel-\sigma_\perp=0$.
For these axes the conductivity tensor is diagonal, and therefore,
quite generally,
\begin{equation}
\rho_\parallel-\rho_\perp
=
\frac{1}{\sigma_\parallel}
-
\frac{1}{\sigma_\perp}
=
-\frac{\sigma_\parallel-\sigma_\perp}
{\sigma_\parallel\sigma_\perp}.
\label{eq:rho_sigma_anisotropy}
\end{equation}
Thus, conductivity and resistivity anisotropies have opposite signs
when they are nonzero, while Eq.~\eqref{main-result} gives
$\rho_\parallel=\rho_\perp$ exactly.

\section{Comparison with previous results}
\label{sec:comparison}

The closest comparison is with Raimondi
\emph{et al.}~\cite{Raimondi2001}. Their model contains the same
ingredients: a single parabolic two-dimensional band, Rashba
spin--orbit coupling, an in-plane Zeeman field, scalar short-range
disorder, and impurity vertex corrections. A relaxation-time
Boltzmann approach with a single transport time and bare band
velocities gives a finite anisotropy. In the weak-disorder two-band
regime, their Kubo calculation without vertex corrections reproduces
this result. Upon including the impurity ladder, however, the dressed
retarded--advanced current vertex becomes approximately the parabolic
one at weak fields, and the conductivity anisotropy changes sign.
Thus, the difference is not between Boltzmann and Kubo approaches,
but between the bare-vertex relaxation-time approximation and the
full impurity collision problem.

The present calculation sharpens this result in the leading two-sheet
quasiclassical sector. The on-shell Born relaxation time is
$\tau_0$, independent of the field and spin, while the impurity ladder
cancels the anomalous Rashba velocity and gives the parabolic current
vertex. The diagonal intraband conductivity then reduces to the total
occupied area and is independent of the field. The residual finite anisotropy in the numerical Kubo calculation of
Ref.~\cite{Raimondi2001} is not the leading diagonal two-sheet
quasiclassical conductivity isolated here. Their numerical analysis
does not isolate the leading term in the asymptotic $E_F\tau_0\gg1$ expansion
and includes interband contributions to the conductivity, which cease
to be negligible when the spin splitting is not resolved on the scale
of the disorder broadening, $\Delta\tau_0\lesssim1$. Their AMR develops
its strongest structure near the spin-polarization scale
$B\sim E_F$. Raimondi \emph{et al.} also obtain an
angle-dependent quasiparticle lifetime in the strong-field limit,
where the upper helicity band is depopulated. That result lies outside
the two-sheet regime considered here.

The relation to Inoue \emph{et al.}~\cite{Inoue2006} is indirect but
important. Their work concerns the anomalous Hall conductivity of a
spin-polarized Rashba model, not the longitudinal AMR considered here.
Nevertheless, it identifies the same kind of cancellation structure:
for short-range disorder, the intrinsic Hall response is removed by the
self-consistent vertex correction unless the quasiparticle lifetime is
spin dependent. Kato \emph{et al.}~\cite{Kato2008} formulated the
corresponding criterion for intrinsic AMR. The Ward identity derived here shows that the required spin dependence
of the Born relaxation time is absent in the two-sheet
Rashba--Zeeman model, so that this criterion gives vanishing diagonal AMR.

The numerical results of Wang and Pang~\cite{WangPang2009} and
Wang~\cite{WangPang2010} should be read in a precise sense. They
establish vanishing AMR for point-like nonmagnetic disorder in the
two-band regime. Their plotted quantities test angular anisotropy under
rotation of the magnetization; they do not formulate the stronger
statement derived here, namely that the diagonal two-sheet
quasiclassical conductivity itself is independent of the in-plane
Zeeman field. The present result therefore goes beyond their
delta-disorder AMR cancellation, but is fully consistent with it. 
Those works also identify the mechanisms that evade the cancellation.
For finite-range or remote scalar disorder, the Born self-energy
contains a momentum-dependent kernel rather than the fully integrated
spectral matrix. A quasiparticle then samples the spin spectral weight
of final states with an angle-dependent form factor. The Ward identity
constrains the unweighted spectral integral, not this
momentum-filtered one. Therefore, finite-range or remote impurity
scattering can generate a genuine angle-dependent transport time and
finite AMR. This is precisely the distinction found in
Refs.~\cite{WangPang2009,WangPang2010,WangPang2011}: zero AMR for
point-like nonmagnetic disorder in the two-band regime and nonzero AMR
for long-range disorder. Wang~\cite{WangPang2010} also emphasized that
the short-range cancellation no longer applies when only one helicity
band is occupied, and that step-like structures appear near the
singular magnetization where the density of states changes
nonanalytically. These regimes are outside the two-sheet
quasiclassical sector considered here.

The bilinear-magnetoresistance theory of Vaz
\emph{et al.}~\cite{Vaz2020} is relevant for a different reason. Their
experimental oxide system is more complex than the present model, but their Rashba model without additional orbital bands uses
local scalar disorder, a Born self-energy, and impurity vertex
corrections. It obtains a field-dependent relaxation-time structure
and a quadratic magnetoresistance already for short-range disorder.
The density Ward identity rules out this relaxation-time mechanism in
the two-sheet Rashba--Zeeman model with delta-correlated
disorder. 
A possible origin of such a field dependence is visible in
relaxation-time calculations. If the angular density of states is
evaluated on the field-deformed Fermi contour while the spin-overlap
factor is evaluated on a reference contour, an apparent
field-dependent and anisotropic relaxation time results. Section S2
of the Supplemental Material shows explicitly that the corresponding
on-shell correction to the spin texture cancels this term. The density
of states and the spin projector must be evaluated on the same
helicity Fermi sheet.
This conclusion concerns the microscopic point-like-disorder
mechanism, not the observation of bilinear magnetoresistance in oxide
heterostructures, where additional orbital bands, energy-dependent
Rashba splittings, Lifshitz transitions, and finite-range disorder may
be relevant.

\section{Conclusion}

We have reexamined anisotropic magnetoresistance in a two-dimensional
Rashba system with an in-plane Zeeman field and point-like nonmagnetic
disorder. When both Rashba-split helicity bands are occupied, and their splitting is well resolved compared with the disorder broadening, the leading quasiclassical diagonal conductivity is
independent of the field and therefore has no AMR.
The Born relaxation time is independent of the field and spin, and
the impurity ladder gives the parabolic transport vertex. The
conductivity then reduces to the total occupied area, which is fixed
by the carrier concentration, giving $\sigma=e^2n\tau_0/m$.

This result also identifies how the cancellation can be avoided.
Finite-range disorder probes a momentum-dependent spectral weight
that is not constrained by the density Ward identity, while interband
mixing, a single occupied helicity sheet, additional band structure,
and effects beyond the leading quasiclassical approximation lie
outside the regime considered here. A field-dependent relaxation time
obtained from point-like scalar disorder within the two-sheet on-shell
Born problem is inconsistent with the Ward identity. In this setting,
an experimentally observed finite AMR is therefore a genuine signature
of physics beyond the quasiclassical mechanism for point-like disorder.

 \acknowledgments

 A.~K. gratefully acknowledges the hospitality of the KIT, Karlsruhe, during his
 scientific visit.

\section*{Appendix}
\appendix

\setcounter{equation}{0}

\section{Area identity for the two-sheet Rashba spectrum}
\label{app:dos_independence}

In this appendix, we prove the geometric identity used in
Sec.~\ref{sec:tau_ward_identity}. We first establish it for the
isotropic Rashba model and then derive the corresponding formula for a
general spin-orbit field linear in momentum. The proof is carried out
for a parabolic dispersion with both helicity sheets occupied. No
expansion in the spin-orbit coupling or in the Zeeman field is made.

\subsection{Isotropic Rashba model}
\label{app:isotropic_area}

For the isotropic Rashba model, introduce
\begin{equation}
\lambda=m\alpha,
\quad
R^2=2mE_F-\lambda^2=2m\mu+\lambda^2,
\quad
b=B/\alpha.
\label{eq:area_identity_lambda_condition}
\end{equation}
Thus,
\begin{equation}
p_F^2=R^2+\lambda^2,
\qquad
2m\mu=R^2-\lambda^2.
\label{eq:R_relations}
\end{equation}
At $B=0$, the two Fermi radii are $p_\pm=R\mp\lambda$.
Consequently, $\lambda=R$ is precisely the Rashba degeneracy
(Dirac) point $\mu=0$. We assume
\begin{equation}
B<\mu,
\label{eq:A_twosheet}
\end{equation}
so that the $s=+$ sheet is occupied. This is the two-sheet regime used
throughout the main calculation. In terms of $R$ and $\lambda$, this
condition reads
\[
R^2-\lambda^2>2mB.
\]

We align the $x$ axis with $\vb B$ and write
\begin{equation}
\vb p=p(\cos\theta,\sin\theta).
\end{equation}
The two dispersions are
\begin{equation}
\epsilon_s(p,\theta)
=
\frac{p^2}{2m}
+s\alpha
\sqrt{p^2+2bp\cos\theta+b^2},
\qquad s=\pm .
\label{eq:app_dispersion}
\end{equation}
In the two-sheet regime, each branch has one positive outer Fermi
radius $p_s(\theta)$, as shown below, and the total occupied area is
\begin{equation}
A_{\rm tot}^{(p)}
=
\frac12
\int_0^{2\pi}d\theta\,
\left[
p_+^2(\theta)+p_-^2(\theta)
\right].
\label{eq:area_polar}
\end{equation}

Squaring the Fermi-contour equation
$\epsilon_s(p,\theta)=\mu$ gives
\begin{equation}
p^4
-2\left(R^2+\lambda^2\right)p^2
-8\lambda^2 b\cos\theta\,p
+\left(R^2-\lambda^2\right)^2
-4\lambda^2 b^2
=0 .
\label{eq:quartic_p}
\end{equation}
Before using this quartic equation, we establish the physical root counting and
thereby exclude spurious roots introduced by squaring.

\subsubsection{Physical roots throughout the two-sheet regime}

The $s=+$ branch has exactly one positive root. Its radial dispersion,
\begin{equation}
\epsilon_+(p,\theta)
=
\frac{p^2}{2m}
+
\alpha
\sqrt{p^2+2bp\cos\theta+b^2},
\end{equation}
is strictly convex on every smooth interval:
\begin{equation}
\frac{d^2\epsilon_+}{dp^2}
=
\frac{1}{m}
+
\alpha
\frac{b^2\sin^2\theta}
{\left(p^2+2bp\cos\theta+b^2\right)^{3/2}}
>0 .
\label{eq:plus_convexity}
\end{equation}
For $\theta=\pi$, the function
$\epsilon_+(p,\pi)=p^2/(2m)+\alpha|p-b|$ has a kink at
$p=b$. On each smooth piece its second derivative is $1/m>0$, and the
function is globally convex. Since
\begin{equation}
\epsilon_+(0,\theta)=B<\mu,
\end{equation}
whereas $\epsilon_+(p,\theta)\to+\infty$ for $p\to\infty$, the equation
$\epsilon_+(p,\theta)=\mu$ has exactly one positive solution, denoted
by $p_+(\theta)$.

Consider now the $s=-$ branch. Away from the isolated point where the
local Rashba--Zeeman splitting vanishes, any physical solution satisfies
\begin{equation}
\frac{p^2}{2m}
=
\mu+
\alpha\sqrt{p^2+2bp\cos\theta+b^2}
>\mu,
\label{eq:minus_p_larger_R}
\end{equation}
and hence
\begin{equation}
p>\sqrt{R^2-\lambda^2}.
\label{eq:minus_root_lower_bound}
\end{equation}
Furthermore,
\begin{equation}
\left|
\frac{p+b\cos\theta}
{\sqrt{p^2+2bp\cos\theta+b^2}}
\right|
\leq 1,
\label{eq:radial_norm_bound}
\end{equation}
so that
\begin{equation}
\frac{d\epsilon_-}{dp}
=
\frac{p}{m}
-
\alpha
\frac{p+b\cos\theta}
{\sqrt{p^2+2bp\cos\theta+b^2}}
\geq
\frac{p}{m}-\alpha.
\label{eq:minus_derivative_bound}
\end{equation}

If
\begin{equation}
\lambda<\sqrt{R^2-\lambda^2},
\end{equation}
Eq.~\eqref{eq:minus_root_lower_bound} immediately gives
\begin{equation}
\frac{d\epsilon_-}{dp}
>
\frac{\sqrt{R^2-\lambda^2}-\lambda}{m}
>0.
\label{eq:minus_monotonic_small_lambda}
\end{equation}

It remains to consider
$\lambda\ge\sqrt{R^2-\lambda^2}$. The two-sheet condition
$B<\mu$ is equivalent to
\begin{equation}
2\lambda b=2mB<R^2-\lambda^2.
\label{eq:two_sheet_lambda_b}
\end{equation}
Hence
\begin{equation}
b<
\frac{R^2-\lambda^2}{2\lambda}
\leq
\frac12\sqrt{R^2-\lambda^2}.
\label{eq:b_bound_strong_soc}
\end{equation}
At a physical lower-branch root,
$p>\sqrt{R^2-\lambda^2}>b$, and therefore
\begin{equation}
\sqrt{p^2+2bp\cos\theta+b^2}\geq p-b.
\end{equation}
The on-shell equation can be written as
\begin{equation}
p^2-\left(R^2-\lambda^2\right)
=
2\lambda
\sqrt{p^2+2bp\cos\theta+b^2}.
\label{eq:minus_onshell_squaredifference}
\end{equation}
Consequently,
\begin{equation}
p^2-\left(R^2-\lambda^2\right)
\geq
2\lambda(p-b),
\end{equation}
or
\begin{equation}
(p-\lambda)^2
\geq
R^2-2\lambda b
>
\lambda^2,
\label{eq:p_minus_lambda_bound}
\end{equation}
where Eq.~\eqref{eq:two_sheet_lambda_b} was used in the last step.
Since $p>0$, Eq.~\eqref{eq:p_minus_lambda_bound} implies
\begin{equation}
p>2\lambda.
\label{eq:p_greater_2lambda}
\end{equation}
Equation~\eqref{eq:minus_derivative_bound} then gives
\begin{equation}
\frac{d\epsilon_-}{dp}
>
\frac{2\lambda}{m}-\alpha
=
\alpha
>0.
\label{eq:minus_monotonic_large_lambda}
\end{equation}
Thus the lower branch is strictly increasing at every possible
crossing throughout the two-sheet regime.

At
$p=\sqrt{R^2-\lambda^2}=\sqrt{2m\mu}$ one has
$\epsilon_-(p,\theta)<\mu$ away from the isolated Rashba--Zeeman
degeneracy, while $\epsilon_-(p,\theta)\to+\infty$ as $p\to\infty$.
There is therefore exactly one positive lower-branch root
$p_-(\theta)$. At the isolated degeneracy point where the square root
in Eq.~\eqref{eq:app_dispersion} vanishes, the same root counting
follows directly, or equivalently by continuity from either side. The
diagonal quasiclassical transport calculation excludes the
disorder-broadened neighborhood of this degeneracy, but the geometric
area identity itself remains continuous there.

\subsubsection{Vieta construction for the area identity}

We can now return to the quartic
Eq.~\eqref{eq:quartic_p}. Substituting $p\to-p$ changes only the sign
of its linear term. This is equivalent to
$\cos\theta\to-\cos\theta=\cos(\theta+\pi)$. Therefore, if
$p_s(\theta+\pi)$ is a positive physical root at angle $\theta+\pi$,
then $-p_s(\theta+\pi)$ is a root of the quartic at angle $\theta$.
The four roots of Eq.~\eqref{eq:quartic_p}, counting multiplicities,
are consequently
\begin{equation}
p_+(\theta),\quad
p_-(\theta),\quad
-p_+(\theta+\pi),\quad
-p_-(\theta+\pi).
\label{eq:quartic_roots}
\end{equation}
This also shows explicitly that squaring the contour equation has not
introduced any additional roots relevant to the Vieta construction:
the two positive roots are the physical roots of the original
$s=\pm$ equations, the two negative roots follow from the
$\theta\to\theta+\pi$ mapping, and these four roots exhaust the quartic.

Define
\begin{equation}
S(\theta)
=
p_+^2(\theta)+p_-^2(\theta).
\label{eq:S_theta}
\end{equation}
The quartic has no cubic term. If its four roots are $r_i$, then
\begin{equation}
\sum_i r_i=0,
\qquad
\sum_{i<j}r_i r_j
=
-2\left(R^2+\lambda^2\right).
\end{equation}
Using
\begin{equation}
\left(\sum_i r_i\right)^2
=
\sum_i r_i^2
+
2\sum_{i<j}r_i r_j,
\end{equation}
we obtain
\begin{equation}
S(\theta)+S(\theta+\pi)
=
4\left(R^2+\lambda^2\right)
=
4p_F^2
=
8mE_F
=
8\pi n.
\label{eq:S_theta_identity}
\end{equation}
Integration over $\theta$ therefore gives
\begin{equation}
A_{\rm tot}^{(p)}
=
\frac12\int_0^{2\pi}\!\!d\theta\,S(\theta)
=
2\pi\left(R^2+\lambda^2\right)
=
4\pi mE_F
=
4\pi^2 n.
\label{eq:Atot_final}
\end{equation}
This result contains no $b=B/\alpha$. Hence
\begin{equation}
\frac{\partial A_{\rm tot}^{(p)}}{\partial \vb B}=0.
\label{eq:Atot_B_independent}
\end{equation}
The result holds throughout the two-sheet domain
$R^2-\lambda^2>2mB$, see Fig.~\ref{Fig:Spectrum}.

\subsection{General linear spin-orbit tensor}
\label{app:A2}

The area identity has a useful generalization to a spin-orbit field
linear in momentum. Consider
\begin{equation}
\hat H(\vb p)
=
\frac{p^2}{2m}\hat{\mathds{I}}
+
\hat{\boldsymbol{\sigma}}\cdot
\left(\mathsf A\,\vb p+\vb B\right),
\label{eq:general_linear_soc_hamiltonian}
\end{equation}
where $\mathsf A$ is a constant in-plane tensor. The helicity
dispersions are
\begin{equation}
\epsilon_s(\vb p)
=
\frac{p^2}{2m}
+
s\left|\mathsf A\,\vb p+\vb B\right|,
\qquad s=\pm.
\end{equation}
For $\vb p=p\vb e_\varphi$, define
\[
a(\varphi)=|\mathsf A\vb e_\varphi|^2,
\qquad
c(\varphi)=\vb B\cdot\mathsf A\vb e_\varphi .
\]
Squaring the Fermi-contour equation gives
\begin{equation}
p^4
-
\left[4m\mu+4m^2a(\varphi)\right]p^2
-
8m^2c(\varphi)p
+
4m^2(\mu^2-B^2)
\!=\!
0.
\label{eq:quartic_general_A}
\end{equation}
We assume that each helicity branch has one positive outer Fermi
radius $p_s(\varphi)$ for every direction.

Since
$a(\varphi+\pi)=a(\varphi)$ and
$c(\varphi+\pi)=-c(\varphi)$, the four roots of
Eq.~\eqref{eq:quartic_general_A} are
\[
p_+(\varphi),\quad
p_-(\varphi),\quad
-p_+(\varphi+\pi),\quad
-p_-(\varphi+\pi).
\]
Vieta's relation therefore gives
\begin{equation}
S_{\mathsf A}(\varphi)+S_{\mathsf A}(\varphi+\pi)
=
8m\mu+8m^2a(\varphi),
\quad
S_{\mathsf A}=p_+^2+p_-^2 .
\label{eq:S_general_A}
\end{equation}
Integrating over $\varphi$ and using
\[\int_0^{2\pi}d\varphi\,
e_{\varphi,i}e_{\varphi,j}=\pi\delta_{ij},\] 
we obtain
\begin{equation}
A_{\rm tot}(\mu,\mathsf A,\vb B)
=
4\pi m\mu
+
2\pi m^2
\operatorname{tr}\!\left(\mathsf A^T\mathsf A\right).
\label{eq:Atot_general_A}
\end{equation}
Thus,
\begin{equation}
\frac{\partial A_{\rm tot}}{\partial\vb B}=0,
\qquad
\frac{\partial A_{\rm tot}}{\partial\mathsf A_{ji}}
=
4\pi m^2\mathsf A_{ji}.
\label{eq:Atot_general_derivatives}
\end{equation}
These derivatives are used below for the density Ward identity and
the current-vertex identity.

\subsection{Momentum-space displacement symmetry and the Ward identity}

The field independence of the occupied area established for the
isotropic Rashba model in Eq.~\eqref{eq:Atot_final} is the geometric
origin of the density Ward identity used in the main text. Changing
the in-plane Zeeman field displaces the Rashba degeneracy point
$\vb p_*=-\vb B/\alpha$ and deforms the individual helicity contours,
while their total occupied area remains unchanged. This displacement
symmetry is especially transparent in Sec.~S3 of the Supplemental
Material, where one introduces
$\vb q=\vb p+\vb B/\alpha$, so that
$|\alpha\vb p+\vb B|=\alpha q$. In the region where the shifted
contours are star-shaped, the area invariance then follows directly
by differentiation with respect to the displacement. The Vieta
construction of Appendix~\ref{app:isotropic_area} establishes the same field independence
throughout the two-sheet domain without relying on that
parameterization, while Eq.~\eqref{eq:Atot_general_A} of Appendix~\ref{app:A2}
shows that the corresponding independence from $\vb B$ persists for
a general spin-orbit field linear in momentum.

The density Ward identity of Sec.~\ref{sec:tau_ward_identity} is the
infinitesimal consequence of this invariance. The density-area
relation
\begin{equation}
n(\mu,\vb B)
=
\frac{A_{\rm tot}^{(p)}}{(2\pi)^2}
\label{density}
\end{equation}
together with Eq.~\eqref{eq:Atot_B_independent}, or with its general
form in Eq.~\eqref{eq:Atot_general_derivatives}, gives
$\partial n/\partial\vb B=0$. Since
$\partial\epsilon_s/\partial\vb B=s\vb\nu_{\vb p}$,
differentiating the density with respect to $\vb B$ produces the
on-shell spin-weighted integral appearing in
Eq.~\eqref{eq:M_density_Ward}. The area invariance therefore implies
that this spin-vector integral vanishes.
The $\vb B$ derivative is therefore the
momentum-space displacement Ward identity. The same occupied-area formula is differentiated in
Appendix~\ref{sec:vertex} with respect to an auxiliary momentum shift
and to the spin-orbit tensor. These derivatives determine,
respectively, the scalar and spin-dependent parts of the
current-vertex correction.

\section{Velocity operator vertex correction}
\label{sec:vertex}

\begin{figure}[!ht]
\vspace{0pt}\includegraphics[width=0.9\columnwidth]{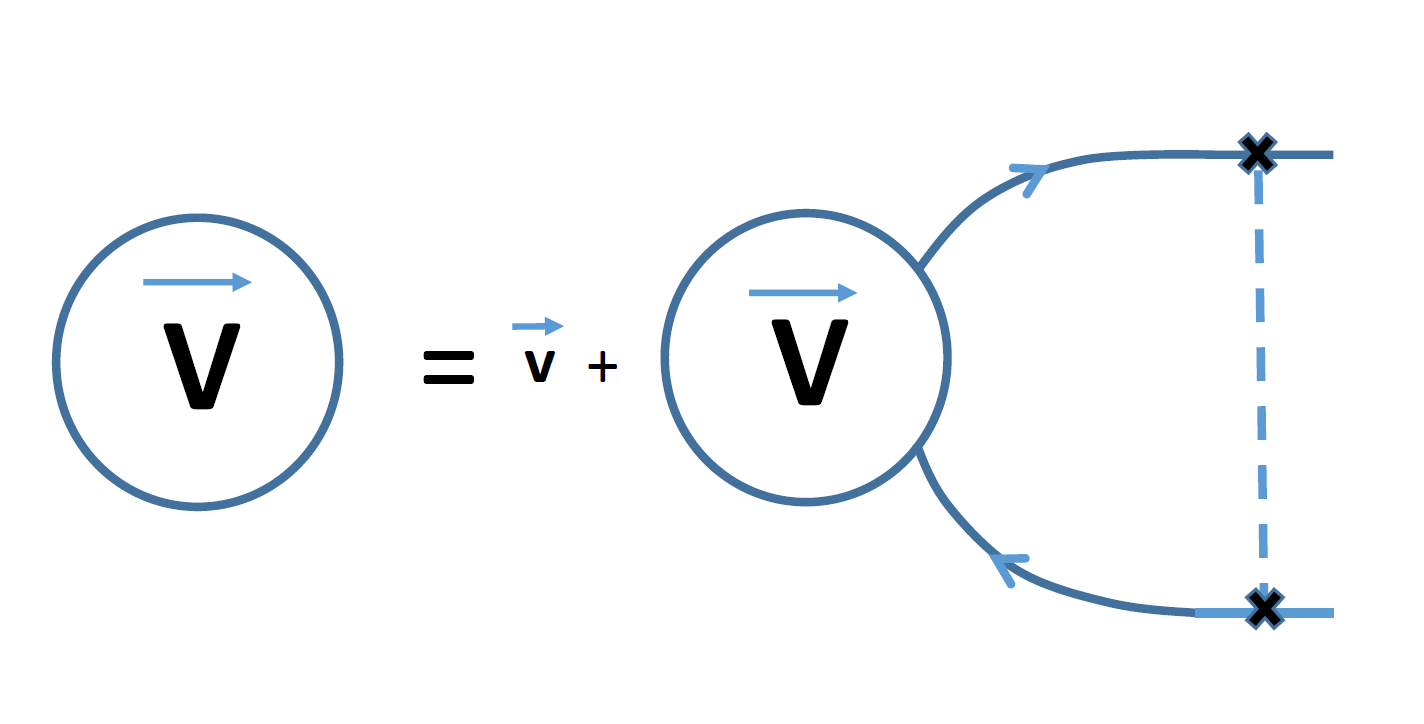}
\caption{\label{Figvertex}
 Diagram for the velocity vertex function. Solid lines are Green's functions; dashed line denotes the impurity correlator. }
\end{figure}

In this appendix, we derive the impurity correction to the velocity vertex in the
same quasiclassical approximation used for the conductivity in the main
text. Both helicity branches are occupied and their local splitting is
resolved compared with the disorder broadening. The calculation shows
how the impurity ladder transforms the bare Rashba velocity into the
transport vertex entering the longitudinal conductivity.
The diagram for the
velocity vertex function is shown in Fig.~\ref{Figvertex}, where the
dotted line between two crosses corresponds to the factor
$1/(m\tau_0)$.

\subsection{Exact quasiclassical ladder solution}
\label{sec:leading}

Here, we derive a current-carrying solution of the
projected retarded--advanced impurity ladder. 
Within the projection specified above, no expansion in $B/E_F$ or
$\alpha p_F/E_F$ is made. Thus, ``exact'' below refers to the projected
retarded--advanced impurity ladder, rather than to finite-broadening or
interband terms in the conductivity. The ratio
$\gamma=B/(\alpha p_F)$ is arbitrary within this regime.

It is useful to prove the vertex identity in a slightly more general
notation. We temporarily write the linear spin-orbit field as
\begin{equation}
 d_j(\vb p)=\mathsf A_{ji}p_i+B_j,
\label{eq:general_d_field}
\end{equation}
where repeated Cartesian indices are summed. The Rashba--Zeeman model
used in the main text is obtained by setting
$\mathsf A_{ji}=\alpha\delta_{ji}$ in the rotated in-plane spin basis
used there. In an unrotated Rashba convention, the same formulas hold
with the corresponding fixed in-plane rotation included in
$\mathsf A$. We also introduce a temporary scalar tilt
$\vb u\cdot\vb p$ and set $\vb u=0$ after differentiating the occupied
area. Thus,
\begin{equation}
 \varepsilon_s(\vb p)
 =
 \frac{p^2}{2m}+\vb u\cdot\vb p+s d(\vb p),
 \quad
 d(\vb p)=|\vb d(\vb p)|,
 \quad
 s=\pm .
\label{eq:general_dispersion_vertex}
\end{equation}
At $\vb u=0$, the two helicity Fermi contours are written as
$p_s(\phi)\vb e_\phi$ and obey
\begin{equation}
 \mu=\frac{p_s^2}{2m}+s d_s,
 \quad
 d_s=|\mathsf A p_s\vb e_\phi+\vb B| .
\label{eq:on_shell_general_vertex}
\end{equation}
We denote the corresponding spin direction by
\begin{equation}
 \nu_{s,j}(\phi)
 =
 \frac{\mathsf A_{ji}p_s e_{\phi,i}+B_j}{d_s},
\label{eq:nu_s_general_vertex}
\end{equation}
and define
\begin{equation}
 D_s(\phi)
 =
 p_s+s m\,\nu_{s,j}\mathsf A_{ji}e_{\phi,i}
 =
 m\frac{\partial\varepsilon_s}{\partial p}
 \bigg|_{p=p_s(\phi),\,\vb u=0} .
\label{eq:D_s_vertex}
\end{equation}

After the radial integration of the retarded--advanced bubble, the
point-disorder ladder equation for the current vertex takes the form
\begin{align}
&\hat\Gamma_i(\vb p)
 =
 \hat v_i(\vb p)\notag \\
&+
 \sum_{s=\pm}\!\int_0^{2\pi}\!\!\frac{d\phi}{2\pi}
 \frac{p_s}{4D_s}
 \left(1+s\vb\nu_s\cdot\hat{\vb\sigma}\right)
 \hat\Gamma_i(p_s\vb e_\phi)
 \left(1+s\vb\nu_s\cdot\hat{\vb\sigma}\right),
\label{eq:vtx_eq_exact}
\end{align}
where
\begin{equation}
 \hat v_i(\vb p)
 =
 \frac{p_i}{m}\hat{\mathds{I}}+
 \mathsf A_{ji}\hat\sigma_j
\label{eq:bare_velocity_general_vertex}
\end{equation}
is the bare velocity operator. We now show that
\begin{equation}
 \hat\Gamma_i(\vb p)=\frac{p_i}{m}\hat{\mathds{I}}
\label{eq:vtx_exact_ansatz}
\end{equation}
solves Eq.~\eqref{eq:vtx_eq_exact} exactly within the two-sheet diagonal
quasiclassical projection.

Substituting Eq.~\eqref{eq:vtx_exact_ansatz} into the ladder correction
in Eq.~\eqref{eq:vtx_eq_exact} gives
\begin{align}
\hat{\mathcal L}_i
 &=
 \sum_{s=\pm}\int_0^{2\pi}\frac{d\phi}{2\pi}
 \frac{p_s^2 e_{\phi,i}}{2mD_s}
 \left(1+s\vb\nu_s\cdot\hat{\vb\sigma}\right)
 \notag\\
 &=
 {\mathcal L}^{(0)}_i\hat{\mathds{I}}
 +
 {\mathcal L}^{(\sigma)}_{ij}\hat\sigma_j,
\label{eq:L_vertex_decomposition}
\end{align}
with
\begin{align}
 {\mathcal L}^{(0)}_i
 &=
 \sum_{s=\pm}\int_0^{2\pi}\frac{d\phi}{2\pi}
 \frac{p_s^2 e_{\phi,i}}{2mD_s},
\label{eq:L0_vertex}\\
 {\mathcal L}^{(\sigma)}_{ij}
 &=
 \sum_{s=\pm}\int_0^{2\pi}\frac{d\phi}{2\pi}
 \frac{s p_s^2 e_{\phi,i}\nu_{s,j}}{2mD_s} .
\label{eq:Ls_vertex}
\end{align}
The two angular identities needed to evaluate these coefficients
follow from derivatives of the occupied area. Let
\begin{equation}
 A_{\rm tot}(\mu,\mathsf A,\vb B,\vb u)
 = \frac12\sum_{s=\pm}\int_0^{2\pi}d\phi\,p_s^2(\phi)
\label{eq:Atot_general_vertex}
\end{equation}
be the total occupied area of the two helicity sheets. For
$\vb u=0$, the two-sheet continuum area is given by Eq.~\eqref{eq:Atot_general_A}
of Appendix~\ref{app:A2} for a general linear
spin-orbit tensor.

In particular, it is independent of
$\vb B$ but has a fixed quadratic dependence on the spin-orbit tensor.

We also introduce a term $\vb u\cdot\vb p$ and set
$\vb u=0$ after differentiating the occupied area.
First, differentiate the area with respect to $u_i$. This is
equivalent to shifting $\vb q=\vb p+m\vb u$:
\begin{equation}
 A_{\rm tot}(\mu,\mathsf A,\vb B,\vb u)
 =
 A_{\rm tot}\!\Bigl(\mu+mu^2/2,\mathsf A,
 \vb B-m\mathsf A\vb u,0\bigr),
\label{eq:tilt_shift_area_vertex}
\end{equation}
so that the derivative at $\vb u=0$ vanishes. Differentiating the on-shell
condition gives
\begin{equation}
 \frac{\partial p_s}{\partial u_i}\bigg|_{\vb u=0}
 =
 -\frac{m p_s e_{\phi,i}}{D_s},
\label{eq:dp_du_vertex}
\end{equation}
and therefore
\begin{equation}
 0
 =
 \frac{\partial A_{\rm tot}}{\partial u_i}\bigg|_{\vb u=0}
 =
 -m\sum_{s=\pm}\int_0^{2\pi}d\phi\,
 \frac{p_s^2 e_{\phi,i}}{D_s} .
\label{eq:ward_tilt_vertex}
\end{equation}
Equation~\eqref{eq:ward_tilt_vertex} gives
\begin{equation}
 {\mathcal L}^{(0)}_i=0 .
\label{eq:L0_zero_vertex}
\end{equation}

Second, we use the derivative of the occupied area with respect to
$\mathsf A_{ji}$ obtained in
Eq.~\eqref{eq:Atot_general_derivatives}. Differentiating the on-shell
condition Eq.~\eqref{eq:on_shell_general_vertex} at fixed $\mu$ gives
\begin{equation}
 \frac{\partial p_s}{\partial\mathsf A_{ji}}
 =
 -\frac{m s p_s e_{\phi,i}\nu_{s,j}}{D_s},
\label{eq:dp_dA_vertex}
\end{equation}
and therefore
\begin{equation}
 \frac{\partial A_{\rm tot}}{\partial\mathsf A_{ji}}
 =
 -m\sum_{s=\pm}\int_0^{2\pi}d\phi\,
 \frac{s p_s^2 e_{\phi,i}\nu_{s,j}}{D_s}.
\label{eq:dA_dA_onshell_vertex}
\end{equation}
Using Eq.~\eqref{eq:Atot_general_derivatives} and comparing with
Eq.~\eqref{eq:Ls_vertex}, we obtain
\begin{equation}
 {\mathcal L}^{(\sigma)}_{ij}
 =
 -\mathsf A_{ji} .
\label{eq:Ls_minus_A_vertex}
\end{equation}
Thus, the ladder correction generated by the trial vertex
Eq.~\eqref{eq:vtx_exact_ansatz} is
\begin{equation}
 \hat{\mathcal L}_i=-\mathsf A_{ji}\hat\sigma_j .
\label{eq:L_minus_anomalous_vertex}
\end{equation}

It cancels the anomalous part of the bare velocity in
Eq.~\eqref{eq:bare_velocity_general_vertex}. Hence
\begin{equation}
 \hat\Gamma_i(\vb p)=\frac{p_i}{m}\hat{\mathds{I}}
\label{eq:vtx_result}
\end{equation}
is an exact current-carrying solution of the projected diagonal
quasiclassical ladder.

For the Rashba convention of the main text this is the vector statement
\begin{equation}
 \vb V=\frac{\vb p}{m} .
\label{eq:V_final_app}
\end{equation}

\subsection{Completeness of the dressed current vertex}
\label{sec:vtx_extra}

The exact calculation in Appendix~\ref{sec:leading} gives the current-carrying solution
$\vb V=\vb p/m$. We now determine whether a homogeneous addition to
this solution can modify the longitudinal conductivity. The
point-disorder ladder first fixes all dependence on the external
momentum, leaving only momentum-independent matrix structures. We
treat the spin-scalar density zero mode separately and then examine
the remaining Pauli-valued homogeneous solutions.

The first statement follows directly from
Eq.~\eqref{eq:vtx_eq_exact}. For point-like disorder, the entire ladder
term contains only the integration variables $p_s(\phi)\vb e_\phi$ and
is independent of the external momentum $\vb p$. Evaluating the
Bethe--Salpeter equation at two arbitrary external momenta $\vb p$ and
$\vb p'$ therefore gives
\begin{align}
 \hat\Gamma_i(\vb p)-\hat\Gamma_i(\vb p')
 &=
 \hat v_i(\vb p)-\hat v_i(\vb p')
 =
 \frac{p_i-p_i'}{m}\hat{\mathds{I}} .
\label{eq:vertex_external_momentum_difference}
\end{align}
Thus,
\begin{equation}
 \hat\Gamma_i(\vb p)-\frac{p_i}{m}\hat{\mathds{I}}
\label{eq:vertex_momentum_independent_remainder}
\end{equation}
is independent of the external momentum. The point-disorder ladder
therefore cannot generate any additional angular or radial dependence.
In particular, a spin-scalar current-like structure
\begin{equation}
 \delta\Gamma_i^{(0)}(\vb p)
 =
 \frac{p_i}{p}F(p/p_0)\hat{\mathds{I}},
\label{eq:forbidden_scalar_momentum_vertex}
\end{equation}
with any ``intrinsic'' scalar momentum scale $p_0$ (for example, $m\alpha$), is
excluded unless $F=0$.

At zero external frequency and wave vector, the spin-scalar identity
matrix $\hat{\mathds{I}}$ is the conserved density eigenmode of the
homogeneous ladder equation, with eigenvalue one, as required by
particle-number conservation.
Hence, the static Bethe--Salpeter vertex equation~\eqref{eq:vtx_eq_exact} formally admits a
momentum-independent vector admixture
\[
 \delta\hat\Gamma_i=c_i\hat{\mathds{I}},
\]
where $\vb c$ is independent of $\vb p$. Such an in-plane vector is
not excluded by symmetry: in the present problem, the Zeeman field
provides a preferred direction, so that one may in particular have
$\vb c\parallel\vb B$, with a scalar coefficient depending on the
parameters of the problem. 
This zero-mode ambiguity, however, has
identically zero overlap with the uniform conductivity.
Indeed, its contribution to the diagonal quasiclassical Kubo bubble is
proportional to
\begin{align}
 \delta\sigma_{ij}
 &\propto
 c_i\sum_{s=\pm}
 \int\frac{d^2p}{(2\pi)^2}\,
 \delta(\mu-\varepsilon_s)
 v_{s,j}
 \notag\\
 &=
 -c_i\sum_{s=\pm}
 \int\frac{d^2p}{(2\pi)^2}\,
 \partial_{p_j}
 \Theta(\mu-\varepsilon_s)
 =0,
\end{align}
where $v_{s,j}=\partial_{p_j}\varepsilon_s$ and the occupied regions
are bounded. Thus, the conserved scalar mode cannot modify any
component of the uniform conductivity, irrespective of the particular
momentum-independent vector coefficient $c_i$.

It remains to determine
whether a momentum-independent Pauli-valued homogeneous solution can
modify the longitudinal conductivity.

A spin-vector correction to the current vertex has the form
$\delta V_k=b_{kj}\hat\sigma_j$, with the tensor $b_{kj}$ built from the
identity and the preferred in-plane direction $\vb e_B$. Its symmetric
part is spanned by $\delta_{kj}$ and $e_{B,k}e_{B,j}$. The antisymmetric
part $\propto\epsilon_{kj}$ belongs to the transverse sector and does not
contribute to the longitudinal anisotropy $\sigma_{xx}-\sigma_{yy}$; the
mixed symmetric structure $e_{B,k}\hat e_{\perp,j}$, with
$\hat{\vb e}_{\perp}=\hat{\vb z}\times\vb e_B$, is odd under reflection
about $\vb e_B$ and is likewise excluded. The general symmetric
spin-vector correction relevant to the AMR is therefore
\begin{equation}
\delta\vb{V}
 =
 \beta\hat{\vb{\sigma}}
 +
 \rho\,\vb{e}_{B}(\vb{e}_{B}\cdot\hat{\vb{\sigma}}),
\label{eq:dV_ansatz}
\end{equation}
an isotropic Rashba-like term and a field-aligned term.

The homogeneous equation for such an additional vertex is obtained by
subtracting the particular solution Eq.~\eqref{eq:V_final_app} from the
Bethe--Salpeter equation. In the same diagonal quasiclassical sector it
reads
\begin{align}
\delta\vb V
 =
 \sum_{s=\pm}\int_0^{2\pi}\frac{d\phi}{2\pi}
 \frac{p_s}{4D_s}
 \left(1+s\vb\nu_s\cdot\hat{\vb\sigma}\right)
 \delta\vb V
 \left(1+s\vb\nu_s\cdot\hat{\vb\sigma}\right),
\label{eq:dV_eq_exact}
\end{align}
with $p_s$, $D_s$, and $\vb\nu_s$ defined in
Eqs.~\eqref{eq:on_shell_general_vertex}--\eqref{eq:D_s_vertex}. Choosing
axes with $\vb e_B=\hat{\vb x}$, the transverse component is
$\delta V_y=\beta\hat\sigma_y$. Projecting the $y$ component of
Eq.~\eqref{eq:dV_eq_exact} onto $\hat\sigma_y$ gives
\begin{equation}
 \beta\left[1-
 \sum_{s=\pm}\int_0^{2\pi}\frac{d\phi}{2\pi}
 \frac{p_s}{2D_s}
 \nu_{s,y}^2
 \right]=0 .
\label{eq:beta_exact_homogeneous}
\end{equation}
Write 
\[\Sigma_a=\sum_{s}\int_0^{2\pi}\frac{d\phi}{2\pi}\frac{p_s}{2D_s}\nu_{s,a}^2\]
for $a=x,y$, so the bracket in Eq.~\eqref{eq:beta_exact_homogeneous} is
$1-\Sigma_y$. Since $\nu_{s,x}^2+\nu_{s,y}^2=1$,
\begin{equation}
\Sigma_x+\Sigma_y
=\sum_{s}\int_0^{2\pi}\frac{d\phi}{2\pi}\frac{p_s}{2D_s}
=\frac{\pi}{m}\frac{\partial n}{\partial\mu}=1 ,
\label{eq:sum_rule_vertex}
\end{equation}
where the last step uses $n=m\mu/\pi+m^2\alpha^2/\pi$ from the area
identity. The bracket therefore equals $\Sigma_x$, which is strictly
positive (the integrand is non-negative and not identically zero), and
$\beta=0$. With $\beta=0$, the field-aligned
component is $\delta V_x=\rho\hat\sigma_x$. Projection of the $x$
component onto $\hat\sigma_x$ gives the analogous equation
\begin{equation}
 \rho\left[1-
 \sum_{s=\pm}\int_0^{2\pi}\frac{d\phi}{2\pi}
 \frac{p_s}{2D_s}
 \nu_{s,x}^2
 \right]=0,
\label{eq:rho_exact_homogeneous}
\end{equation}
whose bracket equals $1-\Sigma_x=\Sigma_y>0$ by
Eq.~\eqref{eq:sum_rule_vertex}. Hence $\rho=0$, and $\beta=\rho=0$.

Thus, the homogeneous ladder equation has no additional
spin-vector solution in the symmetric sector relevant to the
longitudinal conductivity. Together with
Eq.~\eqref{eq:vertex_external_momentum_difference}, which excludes
any additional external-momentum dependence, and the result above
that a momentum-independent vector admixture $c_i\hat{\mathds{I}}$ has
identically zero overlap with the uniform conductivity, this
establishes $\vb p/m$ as the complete transport vertex entering the
projected longitudinal conductivity. The absence of AMR is therefore
not a consequence of restricting the vertex ansatz.

\subsection{Why the transport vertex is essential for the area reduction}
\label{sec:transport_vertex_essential}

The lifetime and the current vertex are fixed by different
derivatives of the same occupied-area formula. Differentiation with
respect to $\vb B$ gives the density Ward identity of
Sec.~\ref{sec:tau_ward_identity} and eliminates the spin-vector part
of the Born broadening. In the ladder calculation, differentiation
with respect to the temporary parameter $\vb u$ in the term
$\vb u\cdot\vb p$ gives ${\mathcal L}^{(0)}_i=0$, while
differentiation with respect to $\mathsf A_{ji}$ gives
${\mathcal L}^{(\sigma)}_{ij}=-\mathsf A_{ji}$. Thus, the scalar
lifetime and the cancellation of the anomalous velocity have the same
geometric origin.

The latter cancellation has a direct transport consequence. Although
the impurity potential is point-like and scalar, projection onto the
helicity states introduces the spinor-overlap factor appearing in
Eq.~\eqref{eq:tau_def_corrected_main}. The resulting transport kernel
is angular dependent, so the dressed current vertex is nontrivial.
The ladder calculation above removes the anomalous spin-orbit velocity,
leaving the parabolic current vertex used in
Sec.~\ref{Sec:Conductivity}. This agrees with the strong ladder
cancellation of the anomalous velocity found by Raimondi
\emph{et al.}~\cite{Raimondi2001}.

This renormalization is what allows the conductivity integral to reduce
to the occupied area. With the dressed current vertex
$\hat V_x=(\vb p/m)_x$ on one side of the retarded--advanced bubble
and the band velocity $v_{s,x}$ on the other side, the branch
integrand becomes Eq.~\eqref{eq:sigma_branch_contour}. The term
containing $\partial_\phi p_s$ then integrates by parts, giving
Eq.~\eqref{eq:sigma_branch_area_form} and finally the area form
Eq.~\eqref{eq:sigma_area_form}. If the anomalous velocity were instead
kept on the dressed leg, so that the bare band velocity $v_{s,x}$ were
used on both sides of the bubble, the integrand would acquire the
additional term
\begin{align}
p_s\,\frac{v_{s,x}^2}{v_{s,\parallel}}
-
p_s\,\frac{p_x\,v_{s,x}}{mv_{s,\parallel}}
&=
s\alpha\,p_s\,\nu_x^{(s)}(\phi)\!
\left(
\cos\phi
+
\frac{\sin\phi}{p_s}\frac{\partial p_s}{\partial \phi}
\right).
\label{eq:anomalous_leftover}
\end{align}
Here, Eq.~\eqref{eq:vx_over_vparallel} has been used and
$\nu_x^{(s)}(\phi)=\vb\nu^{(s)}(\phi)\cdot\vb e_x$. This term is
field dependent through the on-shell spin texture
$\vb\nu^{(s)}(\phi)$ and is not the total-derivative contribution
that produces the geometric area reduction.

Thus, the absence of diagonal AMR in Sec.~\ref{Sec:Conductivity} does
not follow from the isotropy of the bare impurity potential alone.
It relies on the impurity ladder removing the anomalous spin-orbit
velocity from the current vertex, which allows the contour integral
to reduce to the field-independent total occupied area.


\clearpage  
\onecolumngrid  
\begin{center}
  \textbf{\large Supplemental Information}
\end{center}
 
\setcounter{section}{0}
\setcounter{equation}{0}

\renewcommand{\thesection}{S\arabic{section}}
\renewcommand{\thesubsection}{\thesection.\arabic{subsection}}
\renewcommand{\theequation}{S\arabic{equation}}
\makeatletter
\makeatother

This Supplemental Material provides three supporting derivations for the
main text. Section S1 gives the projected Dyson-equation derivation of
the branch-resolved lifetime. Section S2 gives a perturbative weak-field
check showing where an apparent anisotropic scalar lifetime enters and
why it cancels. Section S3 gives an alternative shifted-coordinate proof
of the two-sheet area identity.

\section{Projected Dyson equation and generic branch-dependent lifetime}
\label{app:dyson_details}

This section gives the matrix derivation behind Sec. III of the main text. The main text uses the final scalar result
$\tau_s(\phi)=\tau_0$, but the intermediate steps are useful for two
reasons. First, they show how an angle-dependent lifetime would arise
in a generic spin-orbit model. Second, they make clear why the
short-range Rashba cancellation is nontrivial rather than a consequence
of momentum-independent disorder alone.

\subsection{Born self-energy and radial integral}

The helicity projectors are
\begin{equation}
\hat\Omega_s(\vb p)=\frac{1+s\vb\nu_{\vb p}\cdot\hat{\vb\sigma}}{2},\qquad
\vb\nu_{\vb p}=\frac{\alpha\vb p+\vb B}{|\alpha\vb p+\vb B|} .
\label{eq:S_Omega_def}
\end{equation}
The clean retarded Green's function is
\begin{equation}
\hat G^{R,0}(\epsilon,\vb p)
=
\frac{\hat\Omega_+(\vb p)}
 {\epsilon-\epsilon_+(\vb p)+i0^+}
+
\frac{\hat\Omega_-(\vb p)}
 {\epsilon-\epsilon_-(\vb p)+i0^+}.
\label{eq:G0_app}
\end{equation}
For $\delta$-correlated scalar disorder,
\begin{equation}
\hat\Sigma^R(\epsilon)
=
n_{\rm imp}|U|^2
\int\frac{d^2p'}{(2\pi)^2}\,
\hat G^{R,0}(\epsilon,\vb p').
\label{eq:Sigma_def_app}
\end{equation}
It is independent of the external momentum, but it is a spin matrix.

At fixed angle $\phi'$, the radial integral on branch $s'$ is evaluated
by changing variables to
$\xi=\epsilon_{s'}(p',\phi')-\mu$. On the Fermi surface,
\begin{equation}
p'\,dp'
=
\frac{p_{s'}(\phi')}{v_{s',\parallel}(\phi')}\,d\xi ,
\end{equation}
where
\begin{equation}
v_{s,\parallel}(\phi)
=
\left.
\frac{\partial\epsilon_s(p,\phi)}{\partial p}
\right|_{p=p_s(\phi)},
\qquad
\rho_s(\phi)
=
\frac{p_s(\phi)}{2\pi v_{s,\parallel}(\phi)} .
\label{eq:angDOS_app}
\end{equation}
Extending the $\xi$ integration to the full real axis gives
\begin{align}
\int_0^\infty\frac{p'\,dp'}{2\pi}\,
\frac{\hat\Omega_{s'}(\phi')}
 {\mu-\epsilon_{s'}(p',\phi')+i0^+}
&=
-\rho_{s'}(\phi')
\int_{-\infty}^{\infty}
\frac{d\xi}{\xi-i0^+}\,
\hat\Omega_{s'}(\phi')
\nonumber\\
&=
-i\pi\rho_{s'}(\phi')\hat\Omega_{s'}(\phi')
+
\text{principal value}.
\label{eq:radial_int_app}
\end{align}
The principal value contributes to $\operatorname{Re}\Sigma^R$. The on-shell imaginary part is therefore
\begin{equation}
-\operatorname{Im}\hat\Sigma^R(\mu)
=
\pi n_{\rm imp}|U|^2
\sum_{s=\pm}
\int\frac{d^2p}{(2\pi)^2}
\delta(\mu-\epsilon_s(\vb p))\,
\hat\Omega_s(\vb p).
\label{eq:S_ImSigma_spectral}
\end{equation}

\subsection{Projection of the Dyson equation}

We now project the Dyson equation onto the local helicity basis at the
external momentum $\vb p=p\vb e_\phi$. For any operator $\hat A$,
define
\begin{equation}
A_{ss'}(\phi)
=
\langle s,\phi|\hat A|s',\phi\rangle .
\end{equation}
Although $\hat\Sigma^R$ is momentum independent, its matrix elements
$\Sigma_{ss'}(\phi)$ depend on $\phi$, because the basis states
$|s,\phi\rangle$ rotate with the local spin texture.

The inverse Green's function in this basis is
\begin{equation}
\bigl[(G^R)^{-1}(\phi)\bigr]_{ss'}
=
[\epsilon-\epsilon_s(\phi)]\delta_{ss'}
-
\Sigma_{ss'}(\phi),
\end{equation}
or
\begin{equation}
(G^R)^{-1}
=
\begin{pmatrix}
\epsilon-\epsilon_+(\phi)-\Sigma_{++}(\phi)
&
-\Sigma_{+-}(\phi)
\\
-\Sigma_{-+}(\phi)
&
\epsilon-\epsilon_-(\phi)-\Sigma_{--}(\phi)
\end{pmatrix}.
\label{eq:inverse_G_matrix_app}
\end{equation}
Inverting this matrix gives
\begin{align}
G^R_{ss}(\epsilon,\vb p)
&=
\frac{\epsilon-\epsilon_{-s}(\phi)-\Sigma_{-s,-s}(\phi)}
{\mathcal D(\phi)},
\\
G^R_{s,-s}(\epsilon,\vb p)
&=
\frac{\Sigma_{s,-s}(\phi)}
{\mathcal D(\phi)},
\label{eq:Goffdiag_app}
\end{align}
with
\begin{align}
\mathcal D(\phi)
&=
[\epsilon-\epsilon_+(\phi)-\Sigma_{++}(\phi)]
[\epsilon-\epsilon_-(\phi)-\Sigma_{--}(\phi)]
\nonumber\\
&\quad
-
\Sigma_{+-}(\phi)\Sigma_{-+}(\phi).
\end{align}
Equivalently, the diagonal element may be written as
\begin{equation}
G_s^R(\epsilon,\vb p)
=
\frac{1}
{\epsilon-\epsilon_s(\phi)-\Sigma_{ss}(\phi)
-
\dfrac{\Sigma_{s,-s}(\phi)\Sigma_{-s,s}(\phi)}
{\epsilon-\epsilon_{-s}(\phi)-\Sigma_{-s,-s}(\phi)}} .
\label{eq:full_projection_app}
\end{equation}

This expression separates two issues. The diagonal matrix element
$\Sigma_{ss}(\phi)$ gives the branch-projected broadening and energy
shift. The last term describes the effect of off-diagonal matrix
elements in the local helicity basis.

The momentum independence of $\hat\Sigma^R$ does not imply that
$\Sigma_{ss}(\phi)$ is angle independent. Write
\begin{equation}
\hat\Sigma^R
=
\Sigma_0\hat I+\vb\Sigma\cdot\hat{\vb\sigma}.
\label{eq:Sigma_decomp_app}
\end{equation}
Then,
\begin{equation}
\Sigma_{ss}(\phi)
=
\Sigma_0+s\,\vb\Sigma\cdot\vb\nu^{(s)}(\phi).
\label{eq:Sigma_ss_explicit_app}
\end{equation}
Thus, a fixed spin matrix becomes angle dependent after projection onto a
momentum-dependent helicity basis. In a generic spin-orbit model, a
nonzero spin-vector part of the Born self-energy would produce an
angle-dependent broadening. The Ward identity in Sec. III of the main text shows that this spin-vector part vanishes for the on-shell imaginary self-energy of the continuum short-range Rashba problem.

\subsection{Golden-Rule limit and general branch-resolved rate}

In the clean split-band regime,
\begin{equation}
\Delta(\phi)\tau_0\gg1,
\qquad
\Delta(\phi)=2|\alpha\vb p+\vb B|
\label{eq:clean_limit_app}
\end{equation}
on the Fermi surface, the off-diagonal correction in
Eq.~\eqref{eq:full_projection_app} is small. Since
$\Sigma\sim1/\tau_0$, the last term in Eq.~\eqref{eq:full_projection_app}
is of order $(1/\tau_0)^2/\Delta$, smaller than the diagonal broadening
by $(\Delta\tau_0)^{-1}$. The diagonal Green's function then has the
quasiparticle form
\begin{equation}
G_s^R(\epsilon,\vb p)
\simeq
\frac{1}
{\epsilon-\epsilon_s(\vb p)
-\operatorname{Re}\Sigma_{ss}(\phi)
+i\Gamma_s(\phi)},
\qquad
\Gamma_s(\phi)=-\operatorname{Im}\Sigma_{ss}(\phi).
\label{eq:GR_quasiparticle_app}
\end{equation}

Projecting the on-shell Born spectral matrix gives
\begin{equation}
\frac{1}{\tau_s(\phi)}
=
2\pi n_{\rm imp}|U|^2
\sum_{s'=\pm}
\int_0^{2\pi}\frac{d\phi'}{2\pi}\,
\rho_{s'}(\phi')\,
\frac{
1+ss'\,\vb\nu^{(s)}(\phi)\cdot\vb\nu^{(s')}(\phi')
}{2}.
\label{eq:tau_golden_rule_app}
\end{equation}
The overlap factor follows from
\begin{equation}
\operatorname{tr}\!\left[
\hat\Omega_s(\phi)\hat\Omega_{s'}(\phi')
\right]
=
\frac{
1+ss'\,\vb\nu^{(s)}(\phi)\cdot\vb\nu^{(s')}(\phi')
}{2},
\label{eq:projector_trace_app}
\end{equation}
using $\operatorname{tr}[\hat\sigma_i\hat\sigma_j]=2\delta_{ij}$.

The spin texture in Eq.~\eqref{eq:tau_golden_rule_app} is the on-shell
texture
\begin{equation}
\vb\nu^{(s)}(\phi)
=
\frac{\alpha p_s(\phi)\vb e_\phi+\vb B}
 {|\alpha p_s(\phi)\vb e_\phi+\vb B|}.
\label{eq:nu_s_on_shell_app}
\end{equation}
It should not be replaced by the texture evaluated at $p_F$ unless the
corresponding error is kept consistently.

It is useful to introduce
\begin{equation}
x_s(\phi)=\frac{p_s(\phi)}{p_F},
\qquad
\theta=\phi-\phi_B,
\qquad
\varepsilon=\frac{m\alpha}{p_F},
\qquad
\gamma=\frac{B}{\alpha p_F}.
\end{equation}
As in the main text,
\begin{equation}
E_F=\frac{p_F^2}{2m}=\frac{\pi n}{m},
\qquad
\mu=E_F-m\alpha^2,
\qquad
\frac{2m\mu}{p_F^2}=1-2\varepsilon^2.
\label{eq:SM_mu_EF_relation}
\end{equation}
It is also convenient to denote the field at which the displaced
Rashba degeneracy reaches the Fermi surface by
\begin{equation}
\gamma_{\rm L}
=
\sqrt{\frac{\mu}{E_F}}
=
\sqrt{1-2\varepsilon^2}.
\label{eq:SM_gamma_L}
\end{equation}
\color{black}
Then
\begin{equation}
\vb\nu^{(s)}(\phi)
=
\frac{x_s(\phi)\vb e_\phi+\gamma\vb e_B}
{\sqrt{x_s^2+2\gamma x_s\cos\theta+\gamma^2}}.
\label{eq:nu_s_xs_app}
\end{equation}
The dimensionless on-shell equation is Eq.~\eqref{eq:S_fermi_contour_xs}.
Solving perturbatively gives Eq.~\eqref{eq:S_fermi_contour_expansion}.
The angular density of states is
\begin{equation}
\rho_s(\phi)
=
\frac{m}{2\pi}
\frac{x_s}
{x_s+s\varepsilon\,\vb\nu_{\vb p}\cdot\vb e_\phi}.
\label{eq:rho_xs_app}
\end{equation}

Equations~\eqref{eq:nu_s_xs_app}--\eqref{eq:rho_xs_app} show why the
partially projected expansion is delicate. The branch-dependent shift
of $x_s$ modifies both the density of states and the spin texture. The
terms that appear as an anisotropic density-of-states contribution are
cancelled by the corresponding on-shell correction to $\vb\nu_{\vb p}$. The
full expansion through the order needed for the conductivity calculation
is given in Appendix~\ref{app:tau_expansion_details}. The
nonperturbative Ward identity in Sec. III of the main text explains why
this cancellation persists in the continuum two-sheet short-range
model.

\section{Perturbative check of the scalar Born lifetime}
\label{app:tau_expansion_details}

This section gives a perturbative check of the scalar-lifetime
cancellation derived nonperturbatively in Sec. III and Appendix A of the
main text. The goal is not to rederive the Ward identity, but to show
explicitly how an apparent anisotropic term is removed when all
on-shell quantities are evaluated on the same helicity Fermi sheet.

The branch-resolved rate is
\begin{equation}
\frac{1}{\tau_s(\phi)}
=
2\pi n_{\mathrm{imp}}|U|^2
\sum_{s'=\pm}
\int_0^{2\pi}\frac{d\phi'}{2\pi}\,
\rho_{s'}(\phi')
\frac{
1+s s'\,
\vb\nu^{(s)}(\phi)\cdot\vb\nu^{(s')}(\phi')
}{2}.
\label{eq:tau_def_corrected}
\end{equation}
The spin texture in the overlap must be evaluated on the same Fermi
sheet as the density of states:
\begin{equation}
\vb\nu^{(s)}(\phi)
\equiv
\vb\nu_{\vb p}\big|_{\vb p=p_s(\phi)\vb e_\phi}
=
\frac{\alpha p_s(\phi)\,\vb e_\phi+\vb B}
{\left|\alpha p_s(\phi)\,\vb e_\phi+\vb B\right|}.
\label{eq:nu_s_on_shell_def}
\end{equation}
Replacing $p_s(\phi)$ by $p_F$ in
Eq.~\eqref{eq:nu_s_on_shell_def}, while keeping the density of states
on the displaced Fermi surface, produces a spurious anisotropy. This
calculation is useful because it isolates the step at which an
anisotropic scalar lifetime can enter a relaxation-time treatment.

We use
\begin{equation*}
 \varepsilon=\frac{m\alpha}{p_F},
\qquad
\gamma=\frac{B}{\alpha p_F},
\qquad
\theta=\phi-\phi_B,
\qquad
x_s(\phi)=\frac{p_s(\phi)}{p_F},
\end{equation*}
and
\begin{equation*}
l(\theta)=\sqrt{1+2\gamma\cos\theta+\gamma^2}.
\end{equation*}
Using Eq.~\eqref{eq:SM_mu_EF_relation}, the dimensionless on-shell
equation is
\begin{equation}
x_s^2
+
2s\varepsilon
\sqrt{x_s^2+2\gamma x_s\cos\theta+\gamma^2}
=
1-2\varepsilon^2,
\label{eq:S_fermi_contour_xs}
\end{equation}
with the perturbative solution
\begin{equation}
x_s(\theta)
=1
-s\varepsilon
\sqrt{1+2\gamma\cos\theta+\gamma^2}
-
\frac{\varepsilon^2}{2}(1+\gamma^2)
+
\mathcal O(\varepsilon^3).
\label{eq:S_fermi_contour_expansion}
\end{equation}
\color{black}

The angular density of states is written as
\begin{equation}
\rho_s(\phi)
=
\frac{m}{2\pi}\,
\varrho_s(\phi),
\qquad
\varrho_s(\phi)
=
\frac{x_s}
{x_s+s\varepsilon\,\vb\nu^{(s)}\cdot\vb e_\phi},
\label{eq:rho_varrho_def}
\end{equation}
where the outer Fermi radius is understood. For this root
$v_{s,\parallel}>0$, so the absolute value in the general radial
Jacobian is immaterial.

Using $1/\tau_0=n_{\mathrm{imp}}|U|^2m$,
Eq.~\eqref{eq:tau_def_corrected} becomes
\begin{equation}
\frac{1}{\tau_s(\phi)}
=
\frac{1}{\tau_0}
\left[
\frac{\mathcal S}{2}
+
\frac{s}{2}\,
\vb\nu^{(s)}(\phi)\cdot\vb M
\right],
\label{eq:tau_S_M_form}
\end{equation}
where
\begin{equation}
\mathcal S
=
\sum_{s'=\pm}
\int_0^{2\pi}\frac{d\phi'}{2\pi}\,
\varrho_{s'}(\phi'),
\qquad
\vb M
=
\sum_{s'=\pm}
s'
\int_0^{2\pi}\frac{d\phi'}{2\pi}\,
\varrho_{s'}(\phi')\,\vb\nu^{(s')}(\phi').
\label{eq:S_M_def_app}
\end{equation}
Thus all branch and angular dependence of the scalar lifetime is
contained in the spin-vector spectral weight $\vb M$. The scalar
spectral weight $\mathcal S$ controls only a common, branch-independent
part of the rate.

\subsection{Frozen-texture check}

We first display the error mode explicitly. Suppose that the angular
density of states is evaluated on the true helicity Fermi sheet, but the
spin projector is evaluated on the reference contour $p=p_F$. This is
not a consistent on-shell projection, but it shows how the apparent
anisotropic lifetime is generated.

To the order needed here,
\begin{equation}
\varrho_s(\phi)
=
1-s\varepsilon
+
\mathcal O(\varepsilon\gamma^2,\varepsilon^2).
\label{eq:varrho_spurious_demo}
\end{equation}
If the spin texture is frozen on the reference contour $p=p_F$, one
uses
\begin{equation}
\vb\nu^{(0)}(\phi)
=
\frac{\vb e_\phi+\gamma\vb e_B}
{\sqrt{1+2\gamma\cos\theta+\gamma^2}}
=
\vb e_\phi+\gamma\vb a(\phi)+\mathcal O(\gamma^2),
\label{eq:nu0_spurious_demo}
\end{equation}
with
\begin{equation}
\vb a(\phi)
=
\vb e_B-\cos\theta\,\vb e_\phi .
\end{equation}
The corresponding spin-vector spectral weight would be
\begin{align}
\vb M_{\rm fr}
&=
\sum_{s=\pm}
s\int_0^{2\pi}\frac{d\phi}{2\pi}\,
\varrho_s(\phi)\,\vb\nu^{(0)}(\phi)
=
\sum_s
s\int_0^{2\pi}\frac{d\phi}{2\pi}
\left(1-s\varepsilon\right)
\left(\vb e_\phi+\gamma\vb a\right)
+\cdots
\nonumber\\
&=
-2\varepsilon\gamma
\int_0^{2\pi}\frac{d\phi}{2\pi}\,\vb a(\phi)
+\cdots
=
-\varepsilon\gamma\,\vb e_B+\cdots ,
\label{eq:M_frozen_texture}
\end{align}
where
\begin{equation}
\int_0^{2\pi}\frac{d\phi}{2\pi}\,\vb a(\phi)
=
\frac12\,\vb e_B .
\end{equation}
Substitution into Eq.~\eqref{eq:tau_S_M_form} would therefore give
\begin{equation}
\left.
\frac{1}{\tau_s(\phi)}
\right|_{\rm fr}
=
\frac{1}{\tau_0}
\left[
1
-
\frac{s\varepsilon\gamma}{2}\cos(\phi-\phi_B)
+\cdots
\right].
\label{eq:tau_spurious_anisotropy}
\end{equation}
This is the spurious branch-dependent harmonic produced by the
mismatched projection.

The missing term is the on-shell correction to the spin texture. From
Eq.~\eqref{eq:S_fermi_contour_xs}, the Fermi-radius displacement is
$\delta x_s=-s\varepsilon l+\mathcal O(\varepsilon^2)$.  One then finds
\begin{equation}
\vb\nu^{(s)}(\phi)
=
\vb\nu^{(0)}(\phi)
+
s\varepsilon\gamma\,\vb a(\phi)
+
\mathcal O(\varepsilon\gamma^2,\varepsilon^2).
\label{eq:delta_nu_on_shell_demo}
\end{equation}
This correction contributes
\begin{align}
\delta\vb M
&=
\sum_s
s\int_0^{2\pi}\frac{d\phi}{2\pi}\,
s\varepsilon\gamma\,\vb a(\phi)
+\cdots
=
2\varepsilon\gamma
\int_0^{2\pi}\frac{d\phi}{2\pi}\,\vb a(\phi)
+\cdots
=
+\varepsilon\gamma\,\vb e_B+\cdots .
\label{eq:M_on_shell_correction}
\end{align}
Equations~\eqref{eq:M_frozen_texture} and
\eqref{eq:M_on_shell_correction} cancel. Thus the apparent anisotropic
lifetime in Eq.~\eqref{eq:tau_spurious_anisotropy} is not physical; it
is removed once the density of states and the spin projector are both
evaluated on the same helicity Fermi sheet. The Ward identity in the
main text is the nonperturbative version of this cancellation.

\subsection{Consistent on-shell expansion}

We now repeat the calculation with the consistent on-shell spin texture.
At small $\gamma$,
\begin{equation}
\vb\nu^{(0)}(\phi)
=
\frac{\vb e_\phi+\gamma\vb e_B}{l(\theta)}
=
\vb e_\phi+\gamma\vb a(\phi)+\mathcal O(\gamma^2),
\qquad
\vb a(\phi)=\vb e_B-\cos\theta\,\vb e_\phi .
\label{eq:nu0_a_app}
\end{equation}
The branch-dependent displacement of the Fermi sheet changes the spin
texture by
\begin{equation}
\delta\vb\nu^{(s)}
=
s\varepsilon\gamma\,\vb a(\phi)
+
\mathcal O(\varepsilon\gamma^2,\varepsilon^2).
\label{eq:delta_nu_s_leading}
\end{equation}
Hence
\begin{equation}
\vb\nu^{(s)}(\phi)
=
\vb e_\phi
+
\gamma\vb a(\phi)
+
s\varepsilon\gamma\vb a(\phi)
+
\mathcal O(\gamma^2,\varepsilon\gamma^2,\varepsilon^2).
\label{eq:nu_s_a_expansion}
\end{equation}
At the same order,
\begin{equation}
\varrho_s
=
1
-
s\varepsilon
\frac{1+\gamma\cos\theta}{l(\theta)}
+
\mathcal O(\varepsilon^2)
=
1-s\varepsilon
+
\mathcal O(\varepsilon\gamma^2,\varepsilon^2),
\label{eq:varrho_small_gamma_first_order}
\end{equation}
where the term linear in $\gamma$ cancels in the radial density of
states.

Substituting Eqs.~\eqref{eq:nu_s_a_expansion} and
\eqref{eq:varrho_small_gamma_first_order} into $\vb M$ gives
\begin{align}
\vb M
&=
\sum_s s
\int\frac{d\phi}{2\pi}
\left[
1-s\varepsilon
\right]
\left[
\vb e_\phi
+\gamma\vb a
+s\varepsilon\gamma\vb a
\right]
+
\mathcal O(\varepsilon\gamma^2,\varepsilon^2)
\nonumber\\
&=
\sum_s
\int\frac{d\phi}{2\pi}
\left[
s\vb e_\phi
+
s\gamma\vb a
+
\varepsilon\gamma\vb a
-
\varepsilon\vb e_\phi
-
\varepsilon\gamma\vb a
\right]
+
\mathcal O(\varepsilon\gamma^2,\varepsilon^2).
\label{eq:M_cancellation_app}
\end{align}
The two $\varepsilon\gamma\vb a$ terms cancel before angular
integration. The remaining terms vanish either in the branch sum or by
$\int(d\phi/2\pi)\,\vb e_\phi=0$. Therefore
\begin{equation}
\vb M
=
0
+
\mathcal O(\varepsilon\gamma^2,\varepsilon^2).
\label{eq:M_zero_eps_gamma}
\end{equation}

\subsection{Scalar spectral weight}

For completeness, we also check the scalar spectral weight
$\mathcal S$ to the same order at which it could affect the common
part of the rate. The first-order term in
Eq.~\eqref{eq:varrho_small_gamma_first_order} is odd in the branch
index and therefore cancels in the sum over $s=\pm$. Hence
\begin{equation}
\mathcal S
=
2+\mathcal O(\varepsilon^2).
\label{eq:S_two_leading}
\end{equation}
The $\varepsilon^2$ term can also be evaluated explicitly. Write
\begin{equation}
\vb\nu^{(s)}\cdot\vb e_\phi
=
n_0+s\varepsilon n_1+\mathcal O(\varepsilon^2),
\qquad
n_0=\frac{1+\gamma\cos\theta}{l}.
\end{equation}
Since
$$
\vb\nu\cdot\vb e_\phi
=
\frac{x+\gamma\cos\theta}
{\sqrt{x^2+2\gamma x\cos\theta+\gamma^2}},
$$
and $x_s=1-s\varepsilon l+\mathcal O(\varepsilon^2)$, one obtains
\begin{equation}
n_1
=
-\frac{\gamma^2\sin^2\theta}{l^2}.
\label{eq:n1_radial_app}
\end{equation}
Using Eq.~\eqref{eq:S_fermi_contour_xs}, the denominator of
$\varrho_s$ is
\begin{equation}
x_s+s\varepsilon\,\vb\nu^{(s)}\cdot\vb e_\phi
=
1+s\varepsilon(n_0-l)
+
\varepsilon^2
\left[
-\frac{1+\gamma^2}{2}+n_1
\right]
+
\mathcal O(\varepsilon^3),
\end{equation}
whereas the numerator is
\begin{equation}
x_s
=
1-s\varepsilon l
-
\frac{\varepsilon^2}{2}(1+\gamma^2)
+
\mathcal O(\varepsilon^3).
\end{equation}
\color{black}

Expanding the ratio gives
\begin{equation}
\varrho_s
=
1
-
s\varepsilon n_0
+
\varepsilon^2
\left[
n_0(n_0-l)-n_1
\right]
+
\mathcal O(\varepsilon^3).
\label{eq:varrho_second_order_general_app}
\end{equation}
The bracket simplifies to
\begin{align}
n_0(n_0-l)-n_1
&=
-\frac{\gamma(1+\gamma\cos\theta)(\cos\theta+\gamma)}
 {l^2}
+
\frac{\gamma^2\sin^2\theta}{l^2}
=
-\gamma\cos\theta .
\end{align}
Therefore
\begin{equation}
\varrho_s(\phi)
=
1
-
s\varepsilon
\frac{1+\gamma\cos\theta}{l(\theta)}
-
\varepsilon^2\gamma\cos\theta
+
\mathcal O(\varepsilon^3).
\label{eq:varrho_second_order_result}
\end{equation}
The first correction is odd in $s$ and cancels in $\mathcal S$. The
$\varepsilon^2$ term is independent of $s$, but its angular average
vanishes:
\begin{equation}
\int_0^{2\pi}\frac{d\phi}{2\pi}\cos(\phi-\phi_B)=0.
\end{equation}
Thus
\begin{equation}
\mathcal S
=
2+\mathcal O(\varepsilon^3)
\label{eq:S_two_to_eps2}
\end{equation}
within the one-contour-per-branch regime.

Equations~\eqref{eq:M_zero_eps_gamma} and~\eqref{eq:S_two_to_eps2}
confirm the mechanism behind the Ward-identity result. The anisotropic
term obtained from the density-of-states correction alone is cancelled
by the on-shell correction to the spin texture, and the scalar spectral
weight gives no field-dependent common correction at the displayed
orders. The exact statement, proved in the main text using the Ward identity and the area theorem, is
\begin{equation}
\frac{1}{\tau_s(\phi)}
=
\frac{1}{\tau_0}.
\end{equation}
Thus the perturbative expansion is a consistency check, not an
independent source of corrections to the scalar lifetime.


\section{Weak-field shifted-coordinate proof of the area identity}

This section gives an alternative derivation of the two-sheet area
identity in coordinates centered at the shifted Rashba degeneracy. In
the weak-field regime where the helicity contours are star-shaped with
respect to this shifted origin, the cancellation between the two sheets
is visible directly in the contour parameterization. The global proof
used in the main text is given in Appendix A there.

We align the $x$-axis with $\vb B$ and introduce
\begin{equation}
\vb q=\vb p+\frac{\vb B}{\alpha}.
\end{equation}
Then
\begin{equation}
|\alpha\vb p+\vb B|=\alpha q,
\end{equation}
while the kinetic term becomes
\begin{equation}
\frac{p^2}{2m}
=
\frac{(\vb q-\vb B/\alpha)^2}{2m}.
\end{equation}
Thus, the energy on helicity sheet $s=\pm$ is
\begin{equation}
\epsilon_s
=
\frac{(\vb q-\vb B/\alpha)^2}{2m}
+
s\alpha q.
\end{equation}
Writing
\begin{equation}
b=\frac{B}{\alpha},
\qquad
\lambda=m\alpha,
\qquad
R^2=2mE_F-\lambda^2=2m\mu+\lambda^2,
\end{equation}
and using polar coordinates around the shifted Rashba point,
\begin{equation}
\vb q=q(\cos\theta,\sin\theta),
\end{equation}
the Fermi-contour equation $\epsilon_s=\mu$ becomes
\begin{equation}
q_s^2-2bq_s\cos\theta+b^2+2s\lambda q_s
=
R^2-\lambda^2.
\label{eq:q_contour_equation}
\end{equation}
For
\begin{equation}
b<\sqrt{R^2-\lambda^2}=\sqrt{2m\mu},
\qquad\text{equivalently}\qquad
\gamma<\gamma_{\rm L},
\label{eq:q_star_shaped_condition}
\end{equation}
the contours are star-shaped with respect to $q=0$. The positive
radial solution is then
\begin{equation}
q_s(\theta)
=
b\cos\theta-s\lambda+D_s(\theta),
\label{eq:q_solution}
\end{equation}
where
\begin{equation}
D_s(\theta)
=
\sqrt{(b\cos\theta-s\lambda)^2
+R^2-\lambda^2-b^2}.
\label{eq:Dq_definition}
\end{equation}
\color{black}
The occupied area enclosed by sheet $s$, measured in the same
$\vb q$-plane, is
\begin{equation}
A_s^{(q)}
=
\frac12
\int_0^{2\pi}d\theta\,q_s^2(\theta),
\end{equation}
and the total area is
\begin{equation}
A_{\rm tot}^{(q)}
=
A_+^{(q)}+A_-^{(q)}
=
\frac12
\sum_{s=\pm}
\int_0^{2\pi}d\theta\,q_s^2(\theta).
\label{eq:Atot_q_definition}
\end{equation}

We now show that
\begin{equation}
\frac{\partial A_{\rm tot}^{(q)}}{\partial b}=0.
\label{eq:Atot_q_claim}
\end{equation}
Differentiating Eq.~\eqref{eq:q_contour_equation} with respect to $b$
at fixed $\theta$, and using
\begin{equation}
q_s+s\lambda-b\cos\theta=D_s,
\end{equation}
gives
\begin{equation}
\frac{\partial q_s}{\partial b}
=
\frac{q_s\cos\theta-b}{D_s}.
\label{eq:dqs_db_q}
\end{equation}
Therefore,
\begin{equation}
\frac{\partial A_{\rm tot}^{(q)}}{\partial b}
=
\sum_{s=\pm}
\int_0^{2\pi}d\theta\,
q_s
\frac{q_s\cos\theta-b}{D_s}.
\label{eq:dAtot_db_q_start}
\end{equation}

From Eq.~\eqref{eq:q_solution},
\begin{align}
q_s\cos\theta-b
&=
(b\cos\theta-s\lambda+D_s)\cos\theta-b
=
D_s\cos\theta
-b\sin^2\theta
-s\lambda\cos\theta .
\end{align}
Thus, the contribution of branch $s$ to
Eq.~\eqref{eq:dAtot_db_q_start} can be written as
\begin{align}
I_s^{(q)}
&\equiv
\int_0^{2\pi}d\theta\,
q_s
\frac{q_s\cos\theta-b}{D_s}
=
\int_0^{2\pi}d\theta\,
\left[
q_s\cos\theta
-
\frac{q_s}{D_s}
\left(
b\sin^2\theta+s\lambda\cos\theta
\right)
\right].
\end{align}
Using
\begin{equation}
\frac{q_s}{D_s}
=
1+\frac{b\cos\theta-s\lambda}{D_s},
\end{equation}
we split this as
\begin{align}
I_s^{(q)}
&=
\int_0^{2\pi}d\theta\,
\left[
q_s\cos\theta
-b\sin^2\theta
-s\lambda\cos\theta
\right]
-
\int_0^{2\pi}d\theta\,
\frac{
\left(b\sin^2\theta+s\lambda\cos\theta\right)
\left(b\cos\theta-s\lambda\right)}
{D_s}.
\label{eq:Is_q_split}
\end{align}

Consider first the nonsingular part. Substituting
$q_s=b\cos\theta-s\lambda+D_s$, we find
\begin{align}
\sum_s
\int_0^{2\pi}\!d\theta\,
\left[
q_s\cos\theta
-b\sin^2\theta
-s\lambda\cos\theta
\right]
&=
\sum_s
\int_0^{2\pi}\!d\theta\,
\left[
b(\cos^2\theta-\sin^2\theta)
+D_s\cos\theta
-2s\lambda\cos\theta
\right].
\end{align}
The first and last terms integrate to zero over a full period. The
remaining term cancels between the two helicity branches because
\begin{equation}
D_-(\theta)=D_+(\theta+\pi),
\qquad
\cos(\theta+\pi)=-\cos\theta.
\end{equation}
Hence,
\begin{equation}
\sum_s
\int_0^{2\pi}d\theta\,
D_s(\theta)\cos\theta=0.
\end{equation}

The second term in Eq.~\eqref{eq:Is_q_split} contains the numerator
\begin{align}
\left(b\sin^2\theta+s\lambda\cos\theta\right)
\left(b\cos\theta-s\lambda\right)
&=
b^2\sin^2\theta\cos\theta
+
bs\lambda\cos2\theta
-
\lambda^2\cos\theta .
\end{align}
Therefore,
\begin{equation}
\frac{\partial A_{\rm tot}^{(q)}}{\partial b}
=
-\sum_s
\int_0^{2\pi}d\theta\,
\frac{
b^2\sin^2\theta\cos\theta
+
bs\lambda\cos2\theta
-
\lambda^2\cos\theta
}
{D_s(\theta)}.
\label{eq:singular_q_sum}
\end{equation}
Under $\theta\to\theta+\pi$, one has
$D_-(\theta)=D_+(\theta+\pi)$. The terms proportional to
$\sin^2\theta\cos\theta$ and $\cos\theta$ change sign and cancel
between $s=+$ and $s=-$. The term proportional to $s\cos2\theta$
has a $\pi$-periodic numerator, so the two denominator-mapped
integrals are equal, while the explicit factor $s$ makes them cancel.
Thus, Eq.~\eqref{eq:singular_q_sum} vanishes.

Combining the two parts gives
\begin{equation}
\frac{\partial A_{\rm tot}^{(q)}}{\partial b}=0,
\qquad
\gamma<\gamma_{\rm L}.
\label{eq:Atot_q_derivative_zero}
\end{equation}
\color{black}
This proves the area identity in the regime where the shifted-coordinate
polar representation is globally valid.

The shifted-coordinate derivation is the weak-field local form of the
global area identity proved in Appendix A of the main text. The latter
uses the original momentum coordinate and applies throughout the
continuum two-sheet regime used for the main conductivity calculation.


\end{document}